\documentclass[letterpaper,12pt,peerreviewca,draftcls]{IEEEtran}
\usepackage{csmMBFeb07,graphicx,url}
\usepackage{amssymb}
\usepackage{amsmath}

\newtheorem{theorem}{Theorem}

\newtheorem{lemma}{Lemma}
\newtheorem{definition}{Definition}

\title{Submodularity in Input Node Selection for
Networked Systems\\
{\Large Efficient  Algorithms for Performance and Controllability}
}
\author{Andrew Clark, Basel Alomair, Linda Bushnell, and Radha Poovendran}

\begin{document}
\maketitle
\CSMsetup

Networked systems are systems of interconnected components, in which the dynamics of each component (often referred to as a node or agent) are influenced by the behavior of neighboring components. Examples of networked systems include biological networks, ranging in scale from the gene interactions within a single cell to food webs of an entire ecosystem, critical infrastructures such as power grids, transportation systems, the Internet, and social networks. The growing importance of such systems has led to an interest in control of networks to ensure performance, stability, robustness, and resilience~\cite{ren2008distributed, mesbahi2010graph, farhangi2010path, barabasi2011network}. Indeed, over the past several decades, a variety of control-theoretic methods have been employed to better understand and control networked systems~\cite{antsaklis2007special}. Moreover, new sub-disciplines of control theory have emerged including networked control systems (where plants, sensors, and actuators are connected by communication networks), and control of cyber-physical systems~\cite{wang2008networked,walsh2002stability,hespanha2007survey}.

One approach to control a networked system is to exert control at a subset of nodes, often referred to as input nodes, leaders, driver nodes, or seed nodes, depending on the application domain~\cite{liu2008controllability, kempe2003maximizing, liu2011controllability}. The remaining nodes in the network can then be steered to reach a desired state by exploiting the network interconnections. This approach provides a scalable alternative to directly supplying an input signal to each network node, and is naturally applicable to diverse problems such as steering a flock of unmanned vehicles  from a formation leader~\cite{hu2010distributed}, targeting a set of genes for drug delivery~\cite{rajapakse2012can}, and selectively advertising towards high-influence individuals in a marketing campaign~\cite{kempe2003maximizing, mossel2010submodularity}. 

Under this control framework, the choice of where to exert control becomes a design consideration alongside the choice of how to apply control at selected nodes. The choice of input nodes has been shown to impact a variety of interrelated system properties, including controllability and observability~\cite{tanner2004controllability,cowan2012nodal,liu2013observability,clark2012controllability}; robustness of the system to noise, failures, and attacks~\cite{patterson2010leader,barooah2007estimation}; rate of convergence to a desired state~\cite{rahmani2009controllability, clark2012mixing,clark2013joint}; and the amount of energy required for control~\cite{tzoumas2015minimal}. Enumerating all possible input sets in order to select an optimal set would, however, be computationally infeasible for large-scale networks. The strict performance, robustness, and controllability requirements of networked systems, together with the computational challenges of combinatorial optimization, have motivated the investigation of mathematical structures that can improve the speed and performance of input selection algorithms.

This article presents submodular optimization approaches for input node selection in networked systems. Submodularity is a property of set functions, analogous to concavity of continuous functions, that enables the development of computationally tractable (polynomial-time) algorithms with provable optimality bounds. The submodular structures discussed in this article can be exploited to develop efficient input selection algorithms~\cite{clark2016book}. This article will describe these structures and the resulting algorithms, as well as discuss open problems, and show the practicality of submodular methods for control of networked systems.

The submodular approach to input selection is divided into two components, \emph{submodularity for performance} and \emph{submodularity for controllability}.  Submodularity for performance refers to the ability of the system to reach the desired operating point in a timely fashion  and in the presence of noise, link and node outages, and adversarial attacks. For such metrics, the submodular structure arise from connections between the network dynamics and diffusion processes on the underlying network. 
 
 Controllability refers to the goal of ensuring that the networked system can be driven to a desired operating point  by controlling the input nodes. This article will show that several aspects of the controllability problem  exhibit submodular structure, including the rank of the controllability Gramian, structural controllability (controllability analysis based on the interconnection structure and physical invariants of the networked system),  and the control effort expended by optimal control. Each of these problem types, however, will require a fundamentally different algorithm, and may exhibit differing optimality guarantees. 

This article is organized as follows. Background on submodularity and matroids is given first. The class of networked systems considered in this article are presented. A submodular approach to optimizing performance, including ensuring robustness to noise and smooth convergence, is presented next. Submodular optimization methods for controllability are discussed, followed by techniques for joint performance and controllability. The article concludes with a discussion of open problems and a summary of results.

\section{Background on Submodularity}
\label{sec:submod}
Submodularity is a property of set functions $f: 2^{V} \rightarrow \mathbb{R}$, which take as input a subset of a finite set $V$ and output a real number.   A function $f$ is \emph{submodular} if, for any sets $S$ and $T$ with $S \subseteq T \subseteq V$, and any element $v \in V \setminus T$, 
\begin{equation}
\label{eq:submod_def}
f(S \cup \{v\}) - f(S) \geq f(T \cup \{v\}) - f(T).
\end{equation}
A function $f$ is \emph{supermodular} if $-f$ is submodular. Eq. (\ref{eq:submod_def}) can be interpreted as a \emph{diminishing returns} property, in which adding the element $v$ to a set $S$ has a larger incremental impact than adding $v$ to a superset $T$. This property is analogous to concavity of continuous functions. Functions that are submodular can be efficiently optimized with provable optimality guarantees under a variety of constraints.

\begin{figure}[!ht]
\centering
\includegraphics[width=4in]{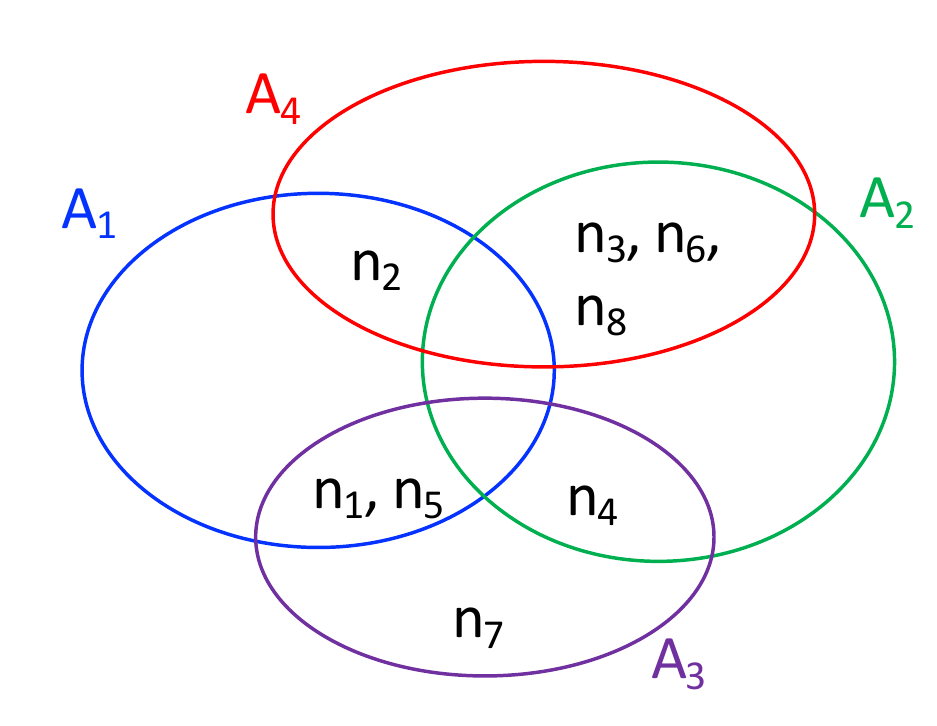}
\caption{Submodularity of the set cover problem. Define $A$ to be a finite set and $R_{1},\ldots,R_{n}$ to be  subsets of $A$. The function $f(S)$ is given by $f(X) = \left|\bigcup_{i \in X}{R_{i}}\right|$, where $| \cdot |$ denotes cardinality of a set. In this example, $A = \{a_{1},\ldots,a_{8}\}$, $R_{1},\ldots,R_{4}$ are defined as shown, $S = \{1\}$, $T = \{1,3,4\}$, and $v=2$. All elements contained in $v$ are already contained in $T$, and hence adding $v$ to $T$ results in no incremental increase in $f$.}
\label{fig:set_cover}
\end{figure}

One example of a problem with submodular structure is set cover, defined as follows. Let $A = \{a_{1},\ldots,a_{m}\}$ denote a finite set, and let $R_{1},\ldots,R_{n}$ denote  subsets of $A$. Define $V = \{1,\ldots,n\}$, and for any subset $X \subseteq V$, let $f(X) = \left|\cup_{i \in X}{R_{i}}\right|$, where $|\cdot|$ denotes the cardinality of a set. The increment $f(S \cup \{v\}) - f(S)$ from adding an element $v$ is equal to the number of elements that are contained in $R_{v}$ but not in $\cup_{i \in S}{R_{i}}$. As the size of $S$ grows, the number of elements in $R_{v} \setminus \cup_{i \in S}{R_{i}}$ decreases (Figure \ref{fig:set_cover}). Hence, the function $f(S)$ is submodular as a function of $S$.

\begin{figure}[!ht]
\centering
\includegraphics[width=5in]{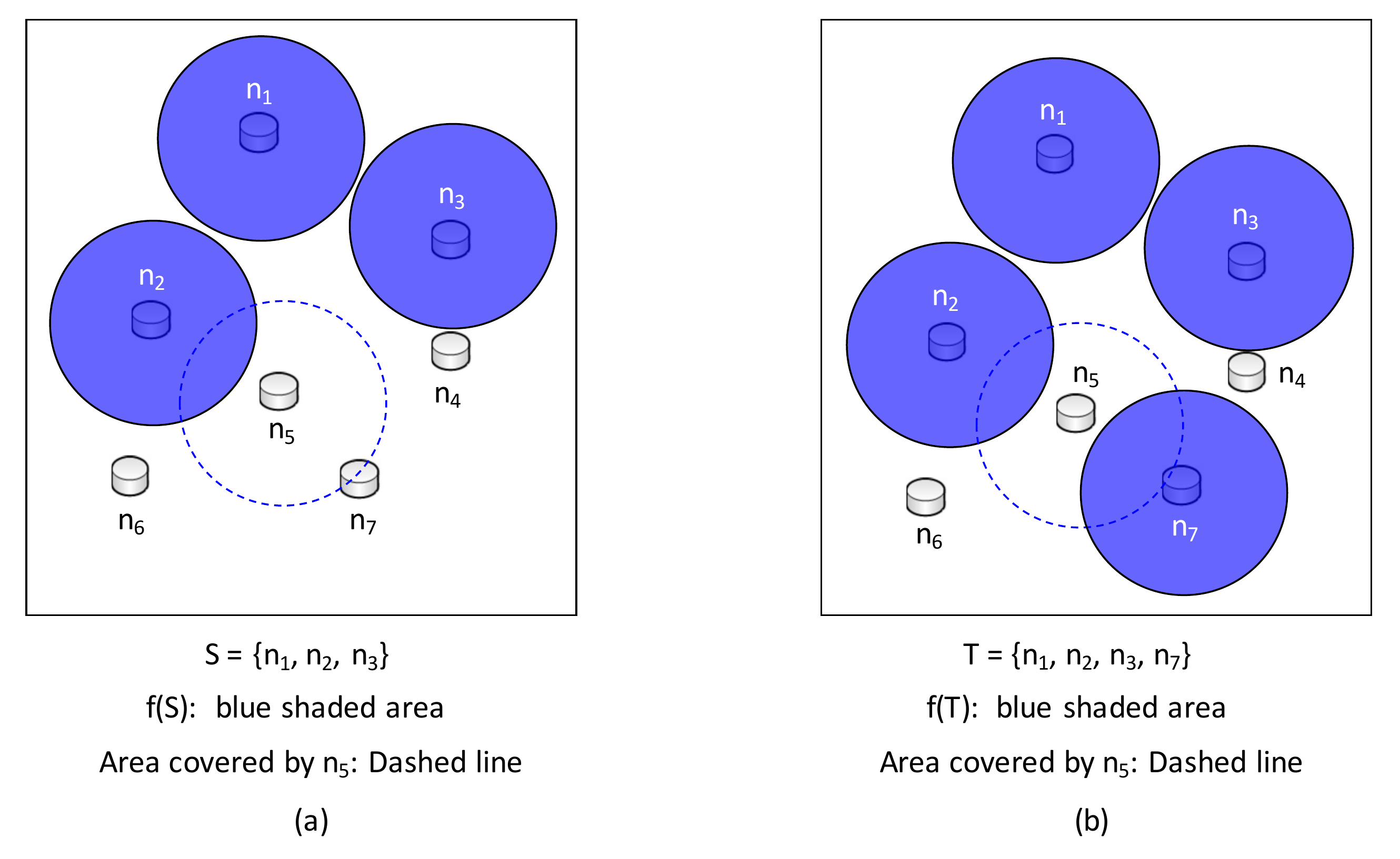}
\caption{Example of submodularity. Each sensor (grey cylinder) has a coverage area, equal to the area that the sensor can monitor. The goal is to monitor the rectangular region by selecting a subset of active sensors. The function $f(S)$ is equal to the total area covered by the set of sensors, denoted $S$. Solid blue circles represent the coverage areas of sensors that are active. The dashed lines indicate the coverage area of the sensor $n_{5}$. (a) Incremental increase in coverage area from adding $n_{5}$ to the set of active sensors $S = \{n_{1}, n_{2}, n_{3}\}$. (b) Incremental increase in coverage area from adding $n_{5}$ to the set $T = \{n_{1}, n_{2}, n_{3}, n_{7}\}$. The incremental increase $f(T \cup \{n_{5}\}) - f(T)$ from adding $n_{5}$ to $T$ is smaller than the incremental increase from adding $n_{5}$ to $S$, since most of $n_{5}$'s coverage area is already covered by the sensors in $T$. Hence $f$ is a submodular function.}
\label{fig:submod}
\end{figure}

An additional example of a problem that exhibits submodularity is shown in Figure \ref{fig:submod}. Each grey cylinder in the figure represents a sensor node, which can monitor a fixed coverage area. A subset of sensors is active at any time, with the number of active sensors limited by battery constraints of each sensor. The goal of covering the entire rectangular region is captured by the function $f(S)$, which is equal to the total area covered by the sensors in set $S$. The incremental increase in coverage area from adding $n_{5}$ to the set of active sensors is larger for a smaller set of active sensors, denoted $S$ (Figure \ref{fig:submod}(a)) than for the larger set $T$ shown in Figure \ref{fig:submod}(b). 

 A function $f: 2^{V} \rightarrow \mathbb{R}$ is monotone nondecreasing (respectively, nonincreasing) if, for any $S \subseteq T$, $f(S) \leq f(T)$ (respectively, $f(S) \geq f(T)$). Not all submodular functions are monotone, and vice versa; however, functions that are both monotone and submodular can be optimized with improved optimality bounds \cite{nemhauser1978analysis,calinescu2011maximizing}.

 Submodular functions have composition rules, analogous to composition of convex functions, that are useful when proving submodularity \cite{fujishige2005submodular}. A nonnegative weighted sum of submodular functions $f_{1}(S),\ldots,f_{m}(S)$ is submodular as a function of $S$. For any monotone submodular function $f(S)$, the function $g(S) = \max{\{f(S),c\}}$ is submodular for any constant $c$. Submodular functions satisfy a complementarity property: if $f: 2^{V} \rightarrow \mathbb{R}$ is a submodular function, then the function $\tilde{f}(S)$ defined by $\tilde{f}(S) = f(V \setminus S)$ is submodular as well. Finally, if $f:2^{V} \rightarrow \mathbb{R}$ is a nonincreasing supermodular function and $g: \mathbb{R} \rightarrow \mathbb{R}$ is an increasing convex function, then the composition $h = g \circ f$ is a supermodular function. 
 
 A variety of problems arising in machine learning, social networking, and game theory have inherent submodular structure, which has led to increased research interest in submodular optimization techniques. For details, see ``Applications of Submodularity''.



\subsection{Matroids}
\label{subsec:matroid}
Matroids can be understood as generalizations of linear systems to discrete set systems. Matroids are sufficiently general to provide insights into graphs, matchings, and linear systems~\cite{oxley2006matroid,welsh2010matroid}. Matroid structure enables efficient solution or approximation of a variety of combinatorial optimization problems. In particular, submodular maximization with matroid constraints is known to have  polynomial-time approximation algorithms \cite{calinescu2011maximizing}.

A matroid $\mathcal{M}$ is defined by an ordered pair $(V, \mathcal{I})$, where $V$ is a finite set and $\mathcal{I}$ is a collection of subsets of $V$.  The collection of  sets $\mathcal{I}$ must satisfy three properties:
\begin{enumerate}
\item[(M1)] $\emptyset \in \mathcal{I}$
\item[(M2)] $Y \in \mathcal{I}$ and $X \subseteq Y$ implies $X \in \mathcal{I}$
\item[(M3)] If $X,Y \in \mathcal{I}$ and $|X| < |Y|$, then there exists $y \in Y \setminus X$ such that $(X \cup \{y\}) \in \mathcal{I}$.
\end{enumerate}
If $X \in \mathcal{I}$, then $X$ is said to be independent in $\mathcal{M}$. 
The three properties of $\mathcal{I}$ can be interpreted by an analogy to linear independence of a collection of vectors (Table \ref{table:matroids}). Let $V$ denote a set of vectors. The empty set of vectors is trivially independent, satisfying the first property. If a set of vectors is linearly independent, then any subset is also independent, thus satisfying the second property. Finally, if two sets of vectors $X$ and $Y$ are linearly independent and $|X| < |Y|$, then there must be at least one element in $Y \setminus X$ that is not in the span of $X$, and hence can be added to $X$ while preserving independence.

\begin{table}[!ht]
\centering
\begin{tabular}{|c|c|c|}
\hline
\textbf{Property} & \textbf{Linear Space} & \textbf{Matroid} \\
\hline
Ground set & Set of vectors $V \subset \mathbb{R}^{n}$ & Finite set $V$ \\
\hline
Independence & Linear independence of a set $X \subset V$ & Set $S \in \mathcal{I}$ \\
\hline
Basis & Linearly independent vectors in $V$ & Maximal independent set \\
 &  that span $V$ & \\
\hline
Rank & Rank(X) = rank of matrix with  & Rank function \\
& column set $X$ & $\rho(X) \triangleq \max{\{|X^{\prime}| : X^{\prime} \subseteq X,  X^{\prime} \in \mathcal{I}\}}$ \\
\hline
\end{tabular}
\caption{Properties of matroids and the analogous properties of linear systems.}
\label{table:matroids}
\end{table}

 A basis is a maximal independent set of the matroid. By property (M3), all bases have the same cardinality; otherwise, if two bases $X$ and $Y$ satisfied $|X| < |Y|$, then an element from $Y$ could be added to $X$ while preserving independence, contradicting maximality of $X$. In the linear independence analogy, the bases correspond exactly to bases of the set of vectors $V$. 

The rank function of a matroid $\mathcal{M}$ is defined by $\rho : 2^{V} \rightarrow \mathbb{Z}_{\geq 0}$, with $$\rho(X) = \max{\{|X^{\prime}| : X^{\prime} \subseteq X \mbox{ and } X^{\prime} \in \mathcal{I}\}}.$$ In words, the rank of $X$ is the maximum-cardinality independent subset of $X$. In the linear independence analogy, the rank of a set of vectors is equivalent to the dimension of the span of those vectors, or the rank of the matrix with those vectors as the columns.

\begin{figure}[!ht]
\centering
\includegraphics[width=4in]{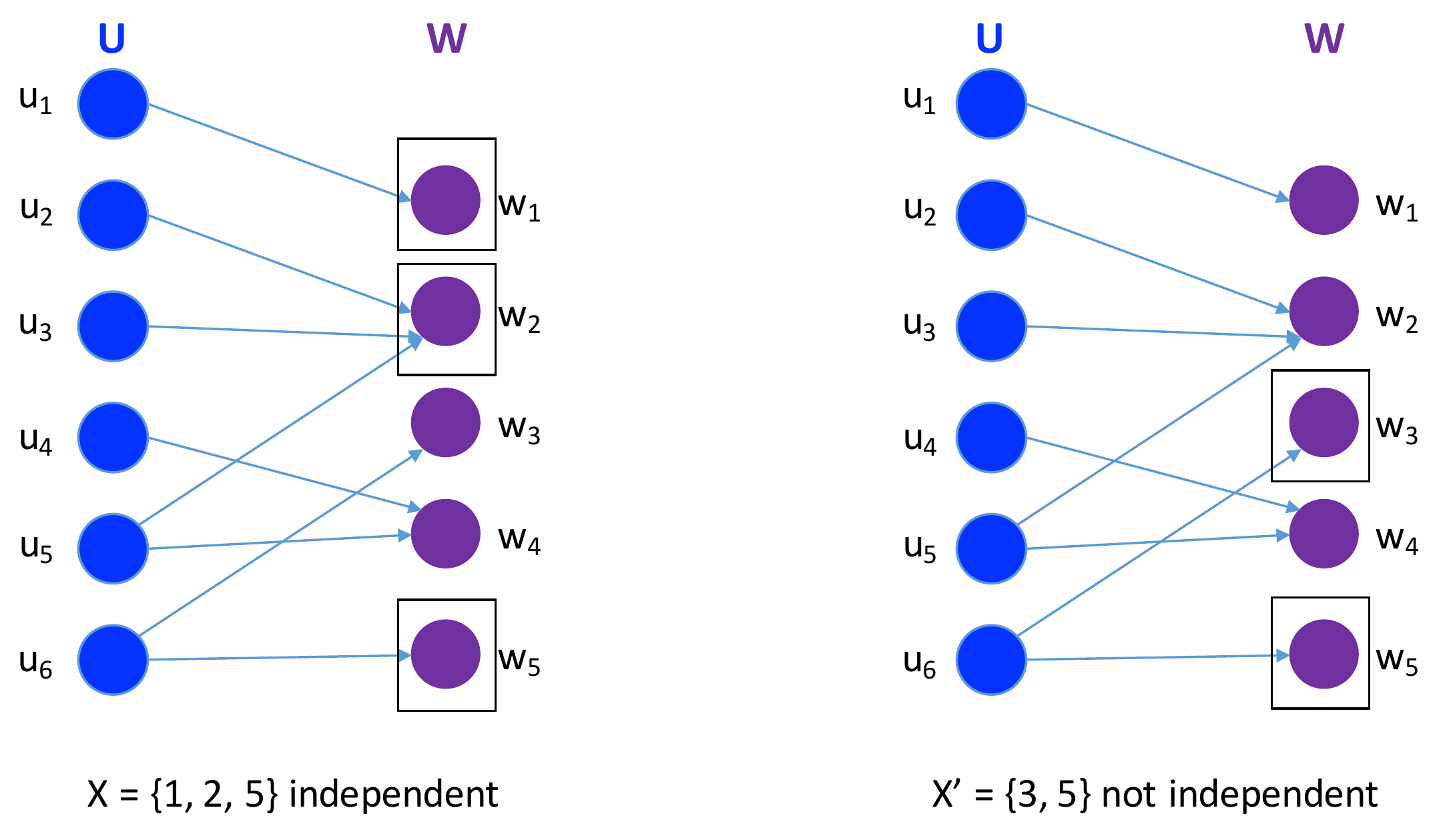}
\caption{A set $X$ is independent in a transversal matroid if there is a matching where each node in $X$ is matched to exactly one node in $W$. In this example, the set $X = \{1, 2, 5\}$ is independent, since there is a matching in which each node in $\{w_{i} : i \in X\}$ is matched to exactly one neighbor in $U$. Note that this matching is not unique: for example, $w_{2}$ could be matched to $u_{2}$, $u_{3}$, or $u_{4}$.  The set $X^{\prime} = \{3, 5\}$ is not independent, since nodes $w_{3}$ and $w_{5}$ can only be matched to the same node  $u_{6}$.} 
\label{fig:transversal}
\end{figure}

A class of matroids that will be useful in the controllability analysis is the \emph{transversal matroids}. Let $U = \{u_{1},\ldots,u_{n}\}$ denote a finite set, and let $W_{1},\ldots, W_{m}$ denote a collection of subsets of $U$. A transversal matroid $\mathcal{M} = (V, \mathcal{I})$ can be defined by setting $V = \{1,\ldots,m\}$ and $$X \in \mathcal{I} \Leftrightarrow \mbox{There exists a one-to-one mapping $f: U \rightarrow X$ with $u_{i} \in W_{f(i)}$}.$$ The transversal matroid can best be interpreted using graph matchings (for details, see ``Graph Matchings''). A bipartite graph can be constructed with vertices indexed $\{u_1,\ldots,u_n\}$ on the left, vertices $\{w_1,\ldots,w_m\}$ on the right, and an edge $(u_i,w_j)$ if $i \in W_{j}$. A set $X \subseteq V$ is independent if there is a matching in this bipartite graph in which $\{w_{i}: i \in X\}$ is matched (Figure \ref{fig:transversal}). 

 \subsection{Submodular Optimization Algorithms}
\label{subsec:algorithms}
Submodular structure enables the development of efficient algorithms with provable optimality guarantees. For a submodular function $f(S)$, two relevant problems that can be formulated within this framework are (a) selecting a set of up to $k$ elements in order to maximize the function $f(S)$, and (b) selecting the minimum-size set of inputs in order to achieve a given bound $\alpha$ on  $f(S)$. The two problems are formulated as
\begin{equation}
\label{eq:opt_setup}
\begin{array}{ccc}
\begin{array}{ll}
\mbox{maximize}_{S \subseteq V} & f(S) \\
\mbox{s.t.} & |S| \leq k
\end{array}
&\quad &
\begin{array}{ll}
\mbox{minimize} & |S| \\
\mbox{s.t.} & f(S) \geq \alpha
\end{array}
\\
(a) & & (b)
\end{array}
\end{equation}

Submodularity implies that simple greedy algorithms are sufficient to approximate both problems up to provable optimality bounds. For Problem \ref{eq:opt_setup}(a), the algorithm is stated as follows:
\begin{enumerate}
\item Initialize the set $S$ to be empty. Set $i=0$.
\item If $i=k$, return $S$. Else go to 3.
\item Select the element $v$ that maximizes $f(S \cup \{v\})$. Update $S$ as $S \leftarrow S \cup \{v\}$. Increment $i$ by 1 and go to 2.
\end{enumerate}
The greedy algorithm returns a set $S^{\prime}$ satisfying $$f(S^{\prime}) \geq \left(1 - \frac{1}{e}\right)f(S^{\ast}),$$ where $S^{\ast}$ is the solution to
(\ref{eq:opt_setup}(a))~\cite{nemhauser1978analysis}.

The algorithm for approximately solving Problem \ref{eq:opt_setup}(b) is similar:
\begin{enumerate}
\item Initialize the set $S$ to be empty. 
\item If $f(S) \geq \alpha$, return $S$. Else go to 3.
\item Select the element $v$ that maximizes $f(S \cup \{v\})$. Update $S$ as $S \leftarrow S \cup \{v\}$. Go to 2.
\end{enumerate}
 Letting $S^{\prime}$ denote the set returned by the second algorithm and $S^{\ast}$ denote the solution to (\ref{eq:opt_setup}(b)), the two sets satisfy $$\frac{|S^{\prime}|}{|S^{\ast}|} \leq 1 + \log{\left\{\frac{f(V)-f(\emptyset)}{f(S^{\prime})-f(S^{\prime\prime})}\right\}},$$ where $S^{\prime\prime}$ denotes the value of the set at the iteration prior to termination of the algorithm~\cite{wolsey1982analysis}. 

Both algorithms terminate in polynomial time and only require $O(n^{2})$ evaluations of the objective function. The greedy algorithm can also be applied to the problem of minimizing a supermodular function to obtain similar guarantees.

\newpage

\section{Sidebar 1: Applications of Submodularity}
\label{subsec:applications}
The intuitive ``diminishing returns'' nature of submodularity leads to inherent submodular structures in a variety of application domains. One such problem consists of selecting a subset of observations to maximize their information content, or equivalently reduce the uncertainty of a random process \cite{krause2008near}\cite{krause2012near}. The submodular structure of this problem arises because many of the classical information-theoretic metrics used to evaluate uncertainty and information gathered, such as entropy and mutual information, are inherently submodular. Practical applications include sensor placement for monitoring temperature, structural health of buildings, and water quality \cite{krause2008near}\cite{krause2008efficient}. Detection of objections in a video frame has also been investigated using submodular optimization methods~\cite{chen2014active}.

\begin{figure}[!ht]
\centering
\includegraphics[width=3in]{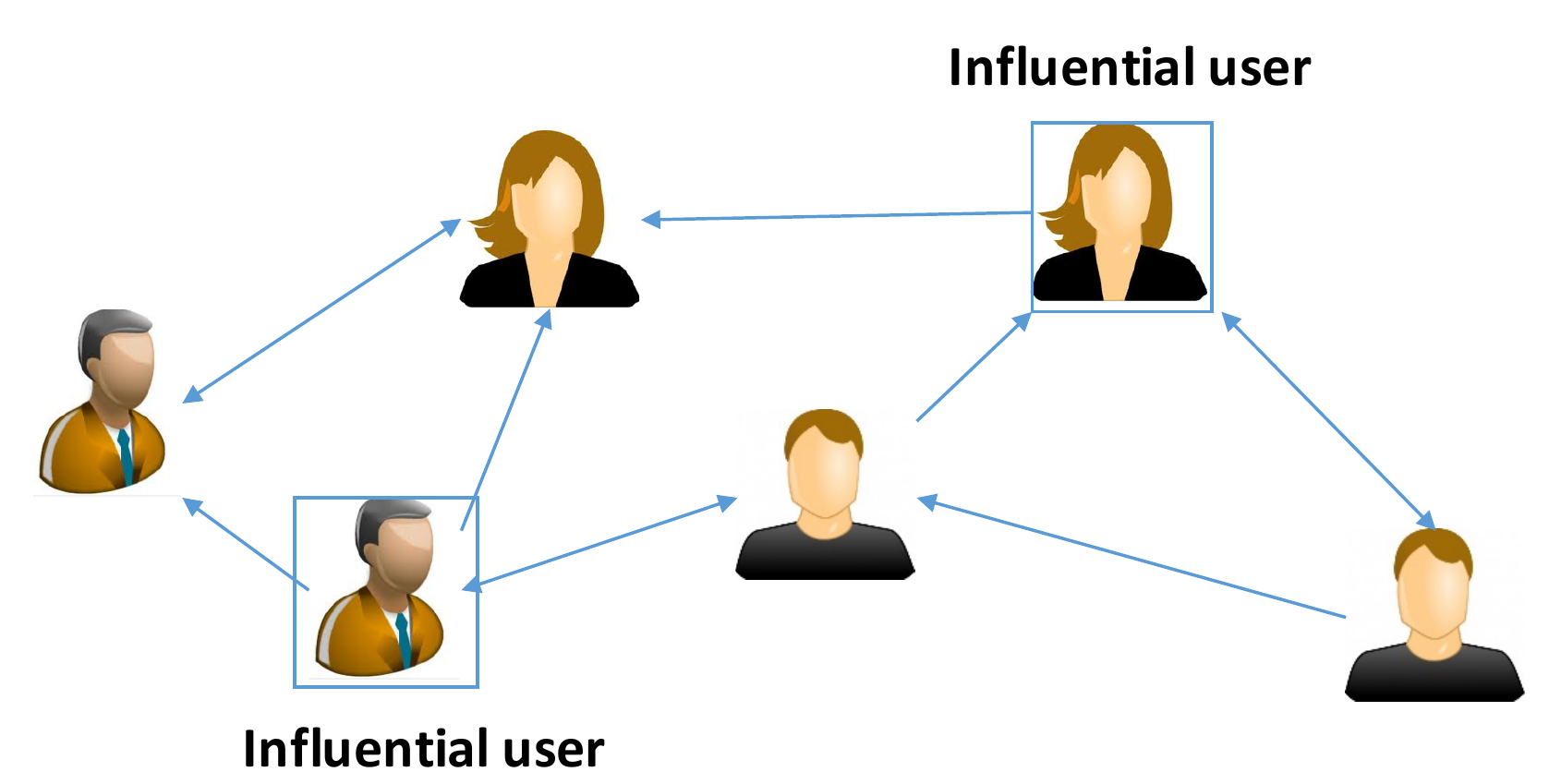}
\caption{Influence maximization problem in social network. The goal of the problem is to select a subset of users, as part of a marketing campaign for example, who are likely to influence other users, creating word of mouth that propagates through the network. The problem of selecting a set of users that maximize the level of influence is known to be submodular for a variety of relevant influence models~\cite{kempe2003maximizing}.}
\label{fig:application_influence}
\end{figure}

Submodular optimization is an important tool in social network analysis, due to the need for scalable algorithms over  network datasets with millions of nodes. A seminal result established that identifying the most influential nodes in a social network can be formulated as a monotone submodular maximization problem \cite{kempe2003maximizing} under several widely accepted influence models (Figure \ref{fig:application_influence}). Related problems, such as selecting sets of influential blogs, have also been studied under the submodular framework~\cite{kimura2007extracting}. An additional problem in social networking consists of identifying the most likely set of links in a social network, based on an observed diffusion process. This learning and estimation problem has also been shown to be submodular \cite{gomez2010inferring}.

Document summarization is the problem of selecting a subset of keywords to best describe a text. Intuitively, the descriptive power of a set of keywords might be expected to be submodular; as more keywords are added to the set, the amount of information gained from each additional keyword diminishes. This intuition was established rigorously by the discovery that some standard information content metrics are submodular as a function of the set of keywords~\cite{lin2011class}. This led to the development of submodular algorithms for keyword selection, which provided better summarization performance than the current state of the art.

Finally, submodularity frequently arises when considering the economic utility of rational entities. Submodularity was proposed in a game-theoretic context  under the class of \emph{convex cooperative games} \cite{shapley1971cores}. In a convex game, a coalition of cooperating players which is stable (no player has an incentive to leave the coalition) can be computed in polynomial time.
\newpage
\section{Sidebar 2: Graph Matching}
\label{sidebar:matching}
 Consider a scenario where $n$ users, denoted $\{u_{1},\ldots,u_{n}\}$, are choosing from a set of $m$ items denoted $\{w_{1},\ldots,w_{m}\}$. Each user $u_{i}$ has a set of ``desirable'' items $W_{i}$. Each user can receive at most one item, and each item can be given to at most one user. The goal in this scenario is to choose a set of (user, item) pairs to maximize the number of users who receive items that are desirable to them.
 
\begin{figure}[!ht]
 \centering
 \includegraphics[width=2in]{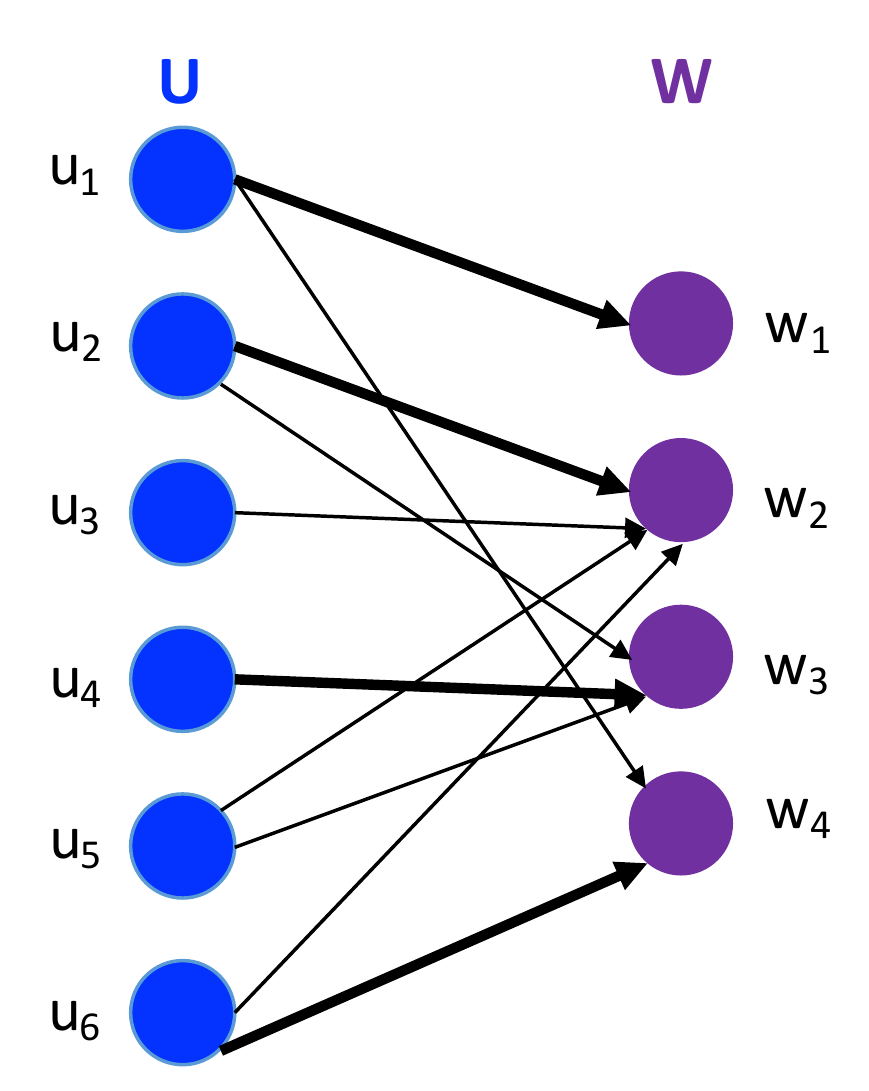}
 \caption{A bipartite graph. Highlighted edges form a maximal matching, in which $u_{1}$ is matched to $w_{1}$, $u_{2}$ is matched to $w_{2}$, $u_{4}$ is matched to $w_{3}$, and $u_{6}$ is matched to $w_{4}$. Note that the matching is not unique, for example, $u_{5}$ could be matched to $w_{3}$ instead.}
 \label{fig:matching}
 \end{figure}

 This problem can be modeled as a \emph{bipartite graph}. A bipartite graph is a graph whose vertex set $V$ can be partitioned into disjoint subsets $U$ and $W$ such that all edges in the graph are between $U$ and $W$ (Figure \ref{fig:matching}). In this scenario, the set $U$ corresponds to users, while the set $W$ corresponds to items, and an edge $(u_{i},w_{j})$ exists if item $w_{j}$ is desirable to user $u_{i}$. 
 
 A \emph{matching} on a bipartite graph is a subgraph in which each node has degree at most one, so that each node in $U$ is ``matched'' to at most one node in $W$. Equivalently, a matching can be viewed as a one-to-one map from $U$ into $W$. A maximal matching is a matching in which no additional edges can be added from $U$ into $W$ while satisfying the requirement that the degree of each node is bounded by one. All maximal matchings can be shown to have the same cardinality \cite{lovasz2009matching}. A matching in which all nodes are matched is a \emph{perfect matching}.
 
 The problem of choosing the mapping between users and items is equivalent to finding a maximal matching on the graph. This problem can be solved in polynomial-time using known methods such as the Edmonds augmenting path algorithm \cite{edmonds1972theoretical}. Other polynomial-time solvable problems include maximum-weight matching (maximizing the utility of the set of users when each user $u_{i}$ attaches benefit $b_{ij}$ to item $w_{j}$) and maximum-weight maximum matching (selecting a maximal matching that maximizes the total benefit) \cite{lovasz2009matching}.
\section{Dynamics of Networked Systems}
\label{sec:dynamics}
A networked system can be modeled as a graph, in which each node of the graph represents one component of the system. A node could represent a single bus in a power system, a gene or protein in a regulatory network, or one vehicle in a transportation network. Each node can be assigned an index $i \in \{1,\ldots,n\}$ in a system of $n$ nodes total. Each node has a time-varying internal state $x_{i}(t) \in \mathbb{R}$. 

The interactions between networked system components are represented by edges in the graph, with two nodes sharing an edge if the dynamics of the corresponding components are coupled to each other. Examples include nodes in communication or formation control networks that are within radio range of each other, and hence can directly share information, buses connected by transmission lines, and genes that directly regulate each other. Together, the nodes and edges form a graph $G=(V,E)$, where $V = \{1,\ldots,n\}$ and $E = \{(i,j): \mbox{$i$ influences the dynamics of $j$}\}$. 

The set of nodes that have an edge incoming to node $i$ is denoted $N_{in}(i) = \{j : (j,i) \in E\}$, and is interpreted as the set of nodes that are influenced by $i$. Similarly, the set of nodes that are influenced by node $i$ is denoted $N_{out}(i) = \{j : (i,j) \in E\}$. The in-degree (respectively, out-degree) is defined by $|N_{in}(i)|$ (respectively, $|N_{out}(i)|$). When the graph is \emph{undirected}, with $(i,j) \in E$ implying $(j,i) \in E$, the in-degree and out-degree are equal and are referred to as the degree $d_{i}$, while $N(i) \triangleq N_{in}(i) = N_{out}(i)$ is the neighbor set of node $i$.

If there is a set of edges $(i_{0},i_{1}), (i_{1},i_{2}), \ldots, (i_{m-1}, i_{m})$ with $i = i_{0}$ and $j = i_{m}$, then this set of edges forms a path from $i$ to $j$. A graph is strongly connected if there is a path between any pair of nodes. 

In some settings, the interactions between neighboring nodes can be treated as a linear coupling, in which the dynamics of node $i$ are given by
\begin{equation}
\label{eq:dynamics}
\dot{x}_{i}(t) = W_{ii}x_{i}(t) + \sum_{j \in N_{in}(i)}{W_{ij}x_{j}(t)}
\end{equation}
where $W_{ij}$ are nonzero weights, when the graph is directed, and
\begin{equation}
\label{eq:undir_dynamics}
\dot{x}_{i}(t) = W_{ii}x_{i}(t) + \sum_{j \in N(i)}{W_{ij}x_{j}(t)}
\end{equation}
when the graph is undirected.

A variety of methods have been developed to control or influence networked systems. These methods include creating or removing links in order to shape the network dynamics \cite{ghosh2006growing}, providing incentives in social networks \cite{marden2014achieving}, and developing distributed strategies for each individual agent to reach a shared goal \cite{leonard2001virtual}. This article considers control techniques in which a subset $S$ of nodes, denoted \emph{input nodes}, have their state values determined directly by an external entity, such as a remote operator, drug intervention, or an external location signal. The states of the input nodes are then treated as external control signals, resulting in a model
$$\dot{\mathbf{x}}_{f}(t) = A_{f}\mathbf{x}_{f}(t) + B_{f}\mathbf{u}(t)$$
where $\mathbf{x}_{f}(t)$ and $\mathbf{u}(t)$ are the state vectors of the non-input and input nodes, respectively. 

The choice of input nodes is known to impact the performance and controllability of networked systems. The impact of the set of input nodes on the robustness of the networked system to noise was studied in \cite{patterson2010leader,barooah2007estimation}. The selected input nodes also determine the rate at which the network dynamics converge to the desired steady-state value \cite{rahmani2009controllability,clark2014minimizing,clark2013joint}. Finally, the controllability of the networked system, defined as the ability of the input nodes to drive the network from any initial state to any final state, depends on which nodes are chosen as inputs \cite{tanner2004controllability,liu2011controllability}. Three input selection problems that can be solved within the submodular optimization framework are:
\begin{enumerate}
\item[Problem 1:] How to select a set of up to $k$ input nodes in order to maximize a performance or controllability metric (analogous to Eq. (\ref{eq:opt_setup}(a)))?
\item[Problem 2:] How to select the minimum-size set of input nodes in order to ensure that the system satisfies a given bound on performance and controllability (analogous to Eq. (\ref{eq:opt_setup}(b)))?
\item[Problem 3:] How to select a set of up to $k$ input nodes in order to maximize a performance metric while guaranteeing controllability?
\end{enumerate}

Heuristics such as selecting high-degree nodes to act as inputs may lead to suboptimal solutions to each of the three problems \cite{clark2014supermodular,liu2011controllability}. The importance of the input nodes motivates the development of an analytical framework for input selection, which will be presented in the following sections.



\newpage
\section{Sidebar 3: Random Walks on Graphs}
\label{subsec:sidebar_RW}
\begin{figure}[!ht]
\centering
\includegraphics[width=3in]{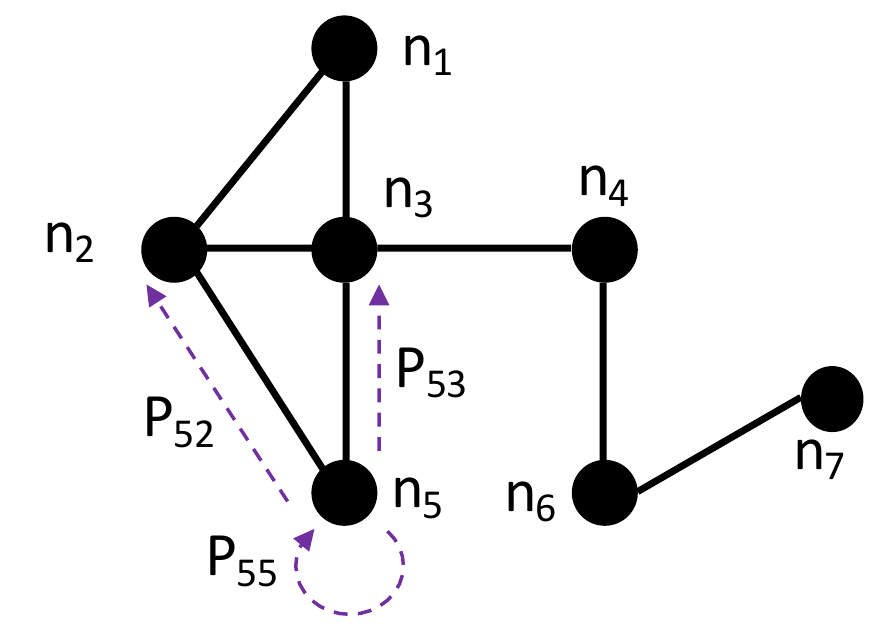}
\caption{Illustration of a random walk with transition matrix $P$ on a graph. $P_{ij}$ denotes the probability of a transition from node $n_{i}$ to node $n_{j}$.}
\label{fig:sidebar_RW}
\end{figure}

A random walk models the behavior of a particle moving at random over a graph $G=(V,E)$ according to a stationary probability distribution.  Formally, a random walk is a discrete-time random process $X[k]$, with $X[k] \in V$.
 A random walk is defined by its transition matrix $P$, where $P_{ij} = P(X[k] = j | X[k-1] = i)$ represents the probability that the walk transitions (takes a step) from location $i$ to location $j$ (Figure \ref{fig:sidebar_RW}). A transition matrix is stochastic, meaning it has nonnegative entries and rows that sum to $1$. 
 
 The behavior of the random walk can be quantified by metrics including the stationary distribution and mixing time. The stationary distribution is the steady-state probability distribution of the walk, and is equal to the solution $\pi$ of $\pi P = \pi$, where $P$ is the transition matrix~\cite{ross1996stochastic}. If the graph is connected and the greatest common divisor of the cycle lengths is 1 (equivalently, the walk is irreducible and aperiodic), then the stationary distribution $\pi$ is unique and the walk converges in probability to the stationary distribution. The mixing time is the rate at which the random walk converges to the stationary distribution, and can be quantified through the eigenvalues of the transition matrix~\cite{levin2009markov}.
 
\begin{figure}
 \centering
 \includegraphics[width=4in]{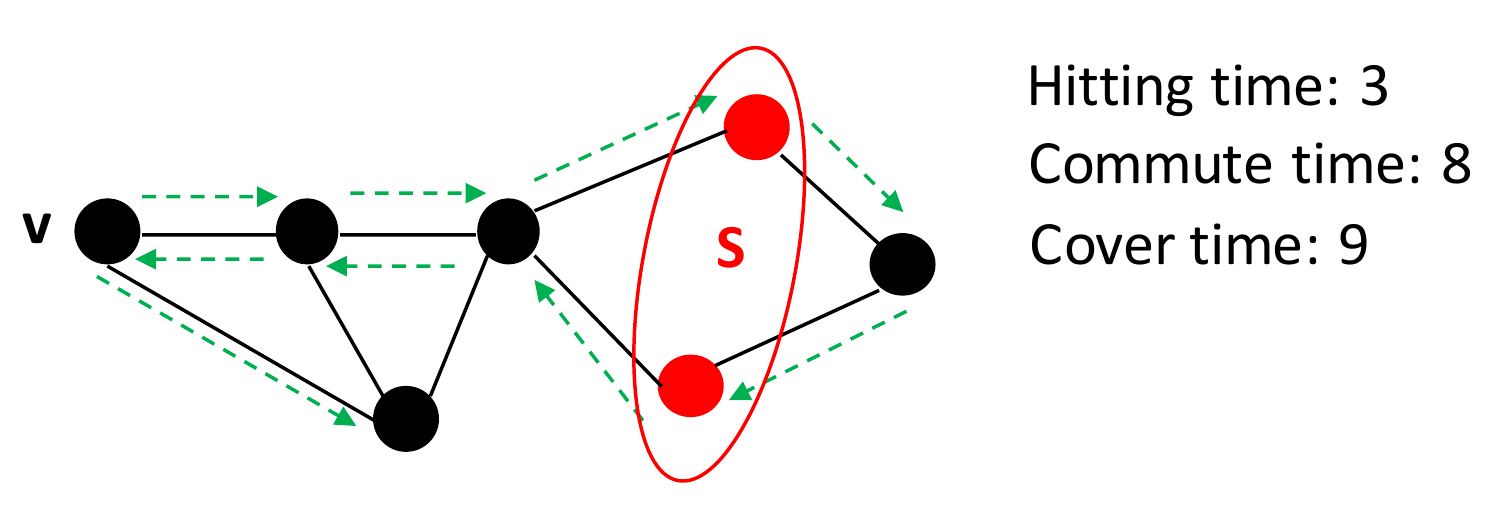}
 \caption{Hitting, commute, and cover times of a random walk on a graph. The times shown represent a single sample path of the walk, indicated by green arrows. The overall hitting, commute, and cover times are obtained by averaging over all sample paths.}
 \label{fig:sidebar_times}
 \end{figure}

 The behavior of a random walk on a graph can be analyzed through statistics including the hitting, commute, and cover times (Figure \ref{fig:sidebar_times}). The hitting time $H(v,S)$ is the expected time for a random walk to reach any node in a set of vertices $S$ from a given initial location $v$.  The commute time $\kappa(v,S)$ is the expected time for a random walk originating at $v$ to reach any node in a set $S$ and then return to $v$. 
 The cover time $C(v)$ is the expected time for a random walk originating at $v$ to reach all vertices in the graph. 
 These statistics are known to have connections to physical quantities including the effective resistance of the graph \cite{doyle1984random}, the rate of diffusion on the graph \cite{lawler2010random}, and the performance of some distributed communication protocols such as gossip and query processing \cite{avin2004efficient, boyd2006randomized}. This article describes the connection between random walks and the performance (robustness to noise and convergence rate) of networked control systems.
\section{Submodularity for Performance: Robustness to Noise}
\label{sec:performance}
 Networked systems operate in inherently lossy and insecure environments, which lead to multiple sources of errors. Noise in communication links can cause incorrect computation of state updates, and hence  error in the state dynamics of the nodes. Physical disturbances are common, as are modeling errors due to simplified models of complex, nonlinear node dynamics and interactions. Errors can also be caused by malicious adversaries through intelligent attacks (such as denial-of service or false data injection). This section considers input selection to minimize the impact of noise for two classes of system dynamics, namely \emph{consensus dynamics} and \emph{Kalman filtering}. Consensus dynamics are widely used in formation control~\cite{ren2008distributed}, distributed estimation~\cite{olfati2005consensus}, and synchronization applications~\cite{he2014time}, while distributed Kalman filtering is a standard approach for estimation and filtering over networks~\cite{roumeliotis2000collective}.

\subsection{Input Selection  for Robustness to Noise}
\label{subsec:robustness}
 Consider a group of unmanned vehicles who exchange information over a communication network. The goal of the vehicles is to follow a trajectory maintained by an external signal (e.g., from a remote operator) sent to the input nodes while maintaining a formation, in a noisy and bandwidth-limited wireless network.
 
 Under the consensus-based approach in the presence of noise,  the input nodes maintain constant states (such as the desired heading or velocity) while the non-input nodes have dynamics 
\begin{equation}
\label{eq:consensus}
\dot{x}_{i}(t) = -\sum_{j \in N(i)}{(x_{i}(t)-x_{j}(t))} + w_{i}(t),
\end{equation}
 where $w_{i}(t)$ is a zero-mean white noise process~\cite{olfati2007consensus, bamieh2012coherence,lin2013leader}. The network topology is assumed to be undirected in this section. An advantage of this approach is that, in the absence of any noise, the states of the nodes will converge to the input node states. Furthermore, the dynamics only require each node to communicate with its one-hop neighbors, reducing the communication overhead required and making the system robust to topology changes. Other applications of the dynamics (\ref{eq:consensus}) include location estimation, time synchronization, and opinion dynamics in social networks. 
In order to further analyze the consensus dynamics with set of input nodes $S$, consider the graph Laplacian matrix defined by 
\begin{equation}
\label{eq:Laplacian}
L_{ij} = \left\{
\begin{array}{ll}
-1, & (i,j) \in E, i \notin S \\
d_{i}, & i=j, i \notin S \\
0, & \mbox{else}
\end{array}
\right.
\end{equation}
The Laplacian matrix can be decomposed as 
\begin{displaymath}
L = \left(
\begin{array}{c|c}
L_{ff} & L_{fl} \\
\hline
0 & 0
\end{array}
\right),
\end{displaymath}
where $L_{ff}$ represents the impact of the non-input nodes on each others' state values, $L_{fl}$ is the impact of the input node states on the non-input nodes, and the remaining zero entries reflect the fact that the input nodes maintain constant state values. The system dynamics can be written in vector form as $$\dot{\mathbf{x}}_{f}(t) = -L_{ff}\mathbf{x}_{f}(t) - L_{fl}\mathbf{x}_{l}(t) + \mathbf{w}(t).$$ 

One metric for the robustness of the system to noise is the $H_{2}$ norm, which is equal to the asymptotic mean-square deviation from the desired state. Letting $\mathbf{x}^{\ast} = \lim_{t \rightarrow \infty}{\mathbf{x}_{f}(t)}$, $\mathbf{x}^{\ast}$ is a random variable with covariance matrix $\Sigma$ given by the solution to the Lyapunov equation $$L_{ff}\Sigma + \Sigma L_{ff} = I.$$ Solving this equation for $\Sigma$ yields $\Sigma=\frac{1}{2}L_{ff}^{-1}$~\cite{patterson2010leader}. The metric $R(S) \triangleq \mathbf{trace}(L_{ff}^{-1})$ therefore quantifies the mean-square error due to noise in the node states. 

 \subsection{Random Walks and Error in Networked Systems}
 \label{subsec:RW_error}
 
 As a first step towards developing a submodular optimization approach to minimizing errors due to noise, a connection will be established between error due to noise and the statistics of a random walk on the network, specifically, the commute time. For details on random walks, see ``Random Walks on Graphs''.
 
 \begin{figure}[!ht]
 \centering
 $\begin{array}{cc}
 \includegraphics[width=3in]{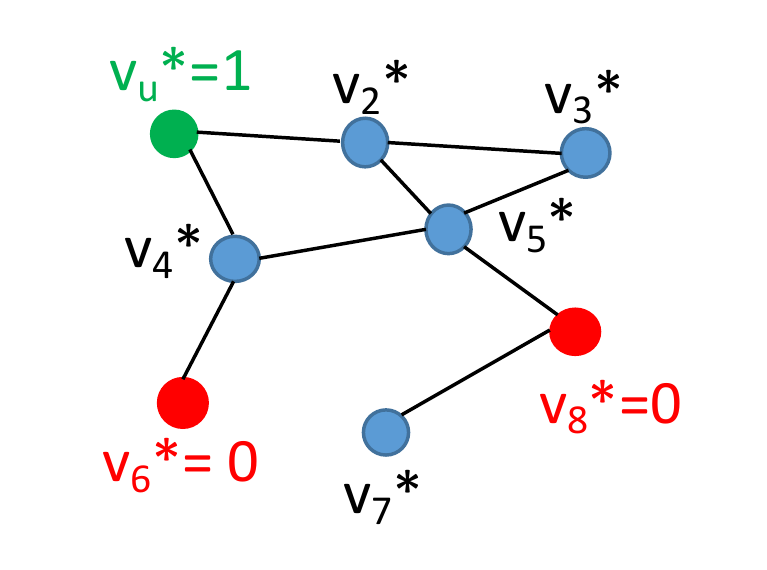} &
 \includegraphics[width=3in]{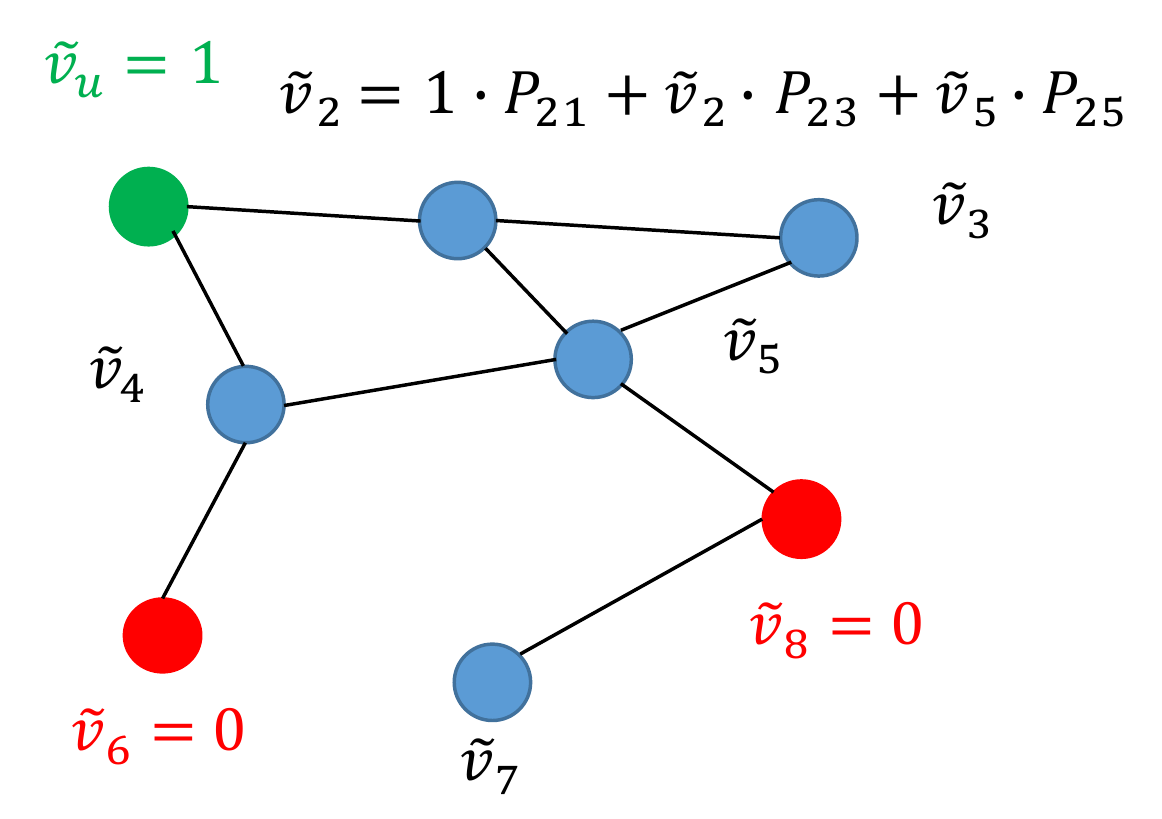} \\
 \mbox{(a)} & \mbox{(b)}
 \end{array}$
 \caption{Description of the connection between the inverse Laplacian and a random walk on the graph. (a) Illustration of the vector $v^{\ast}$ used to compute the error due to noise at node $u$, equal to $(L_{ff}^{-1})_{uu}$,  when $S = \{6,8\}$. (b) Connection to a random walk. The variable $\tilde{v}_{i}$ is equal to the probability that a walk originating at $i$ reaches $u$ before $S$ under a random walk with transition matrix $P$. $P_{21}$, $P_{23}$, and $P_{25}$ denote the transition probabilities of the walk. It can be shown using the Maximum Principle that $v^{\ast} = \tilde{v}$.}
 \label{fig:rw_explanation}
 \end{figure}

 Observe that the $u$-th  diagonal element of $L_{ff}^{-1}$, denoted $(L_{ff}^{-1})_{uu}$, can be obtained as follows. Define vectors $v^{\ast}$ and $J$ as solutions to the equation $Lv^{\ast} = J$ satisfying $v_{u}^{\ast} = 1$, $J_{i} = 0$ for all $i \in V \setminus (S \cup \{u\})$, and $v_{i}^{\ast} = 0$ for all $i \in S$ (Figure \ref{fig:rw_explanation}(a)). Solving this equation for the remaining entires of $v^{\ast}$ and $J$ yields $(L_{ff}^{-1})_{uu} = 1/J_{u}$. Equivalently, $$(L_{ff}^{-1})_{uu} = \left(\sum_{j \in N(u)}{(1-v_{j}^{\ast})}\right)^{-1}.$$ The error variance of node $u$, equal to $(L_{ff}^{-1})_{uu}$, can therefore be expressed as a function of $v^{\ast}$.
 
Note that 
\begin{equation}
\label{eq:walk_intermediate}
v_{i}^{\ast} = \sum_{j \in N(i)}{\frac{|L_{ij}|}{L_{ii}}v_{j}^{\ast}}
\end{equation}
 for all $i \notin (S \cup \{u\})$. On the other hand, consider a random walk with transition matrix $P_{ij} = |L_{ij}|/L_{ii}|$, and define $\tilde{v}_{i}(S,u)$ to be the probability that a random walk originating at $i$ reaches $u$ before $S$ (Figure \ref{fig:rw_explanation}(b)). By construction, $\tilde{v}_{i}(S,u)$ also satisfies (\ref{eq:walk_intermediate}). In fact, it can be shown that $\tilde{v}_{i}(S,u) = v_{i}^{\ast}$ for all $i$ using the maximum principle from harmonic analysis \cite{doyle1984random}, leading to the following result.
 
\begin{theorem}[\cite{clark2014supermodular}]
\label{theorem:commute_time_error}
The variance of the error due to noise at node $u$, $(L_{ff}^{-1})_{uu}$, is proportional to the commute time $\kappa(S,u)$.
\end{theorem}

 
 
 
 
 The connection between the error due to noise and commute time can also be developed using the graph effective resistance. This approach is not presented here due to space constraints, but can be found in \cite{clark2011submodular}. Further reading on the connection between error due to noise and effective resistance can be found in \cite{barooah2007estimation}. 
 



\subsection{Supermodularity of Error Due to Noise}
\label{subsec:RW_supermod}
This subsection establishes supermodularity of the robustness to noise by exploiting the connection to the commute time. The key result is the following.

\begin{theorem}[\cite{clark2014supermodular}]
\label{theorem:commute_supermodular}
The commute time $\kappa(S,u)$ is a supermodular function of the set $S$.
\end{theorem}

As a corollary, the mean-square error metric $R(S)$ is supermodular as well. A sketch of the proof of Theorem \ref{theorem:commute_supermodular} is as follows. First, note that the goal is to show that for any $S \subseteq T$ and any $v \notin T$, $$\kappa(S,u) - \kappa(S \cup \{v\},u) \geq \kappa(T,u) - \kappa(T \cup \{v\},u).$$ Let $U_{jSu}$ be a random variable, equal to the time for a walk starting at a node $j$ to reach $S$ and then reach $u$. Let $\tau_{j}(S)$ denote the event that a walk reaches $j$ before $S$.

The differences in the commute times can then be rewritten using the expression $$\kappa(S,u) - \kappa(S \cup \{j\},u) = \mathbf{E}(U_{jSu}-U_{ju} | \tau_{j}(S))Pr(\tau_{j}(S)),$$ where $\mathbf{E}(\cdot)$ denotes expectation. It therefore suffices to show that $Pr(\tau_{j}(S)) \geq Pr(\tau_{j}(T))$ and $U_{jSu} \geq U_{jTu}$ when $S \cup T$. 

\begin{figure}[!ht]
\centering
$\begin{array}{ccc}
\includegraphics[width=2in]{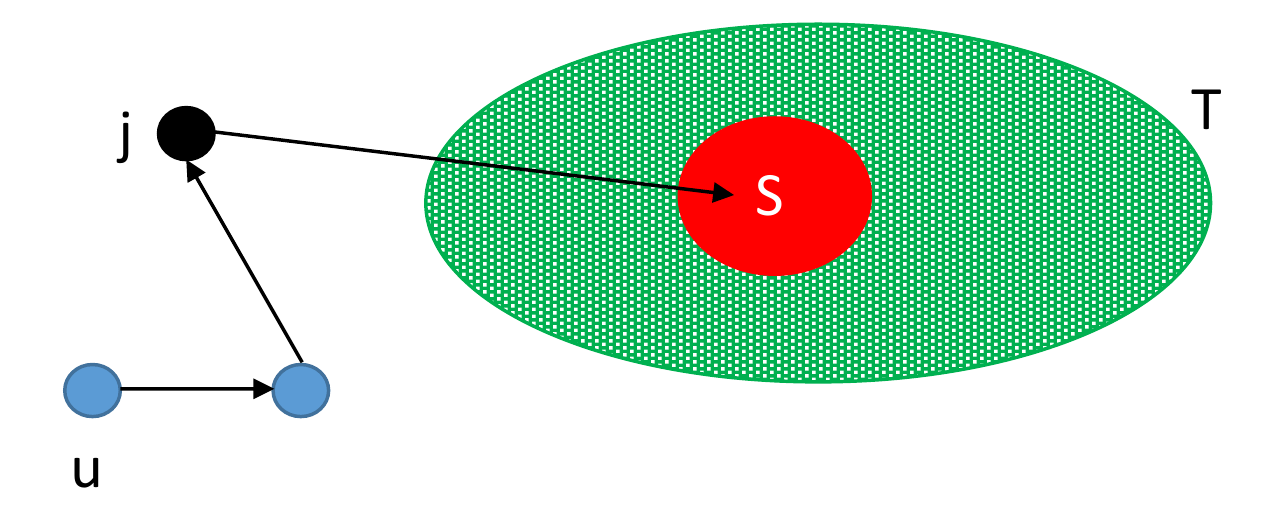} &
\includegraphics[width=2in]{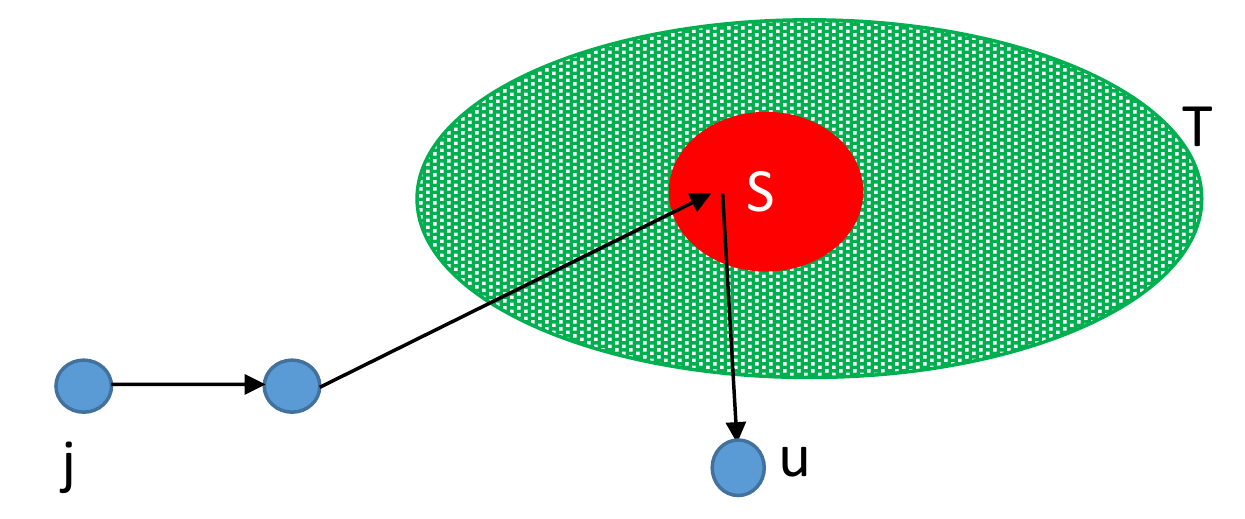} &
\includegraphics[width=2in]{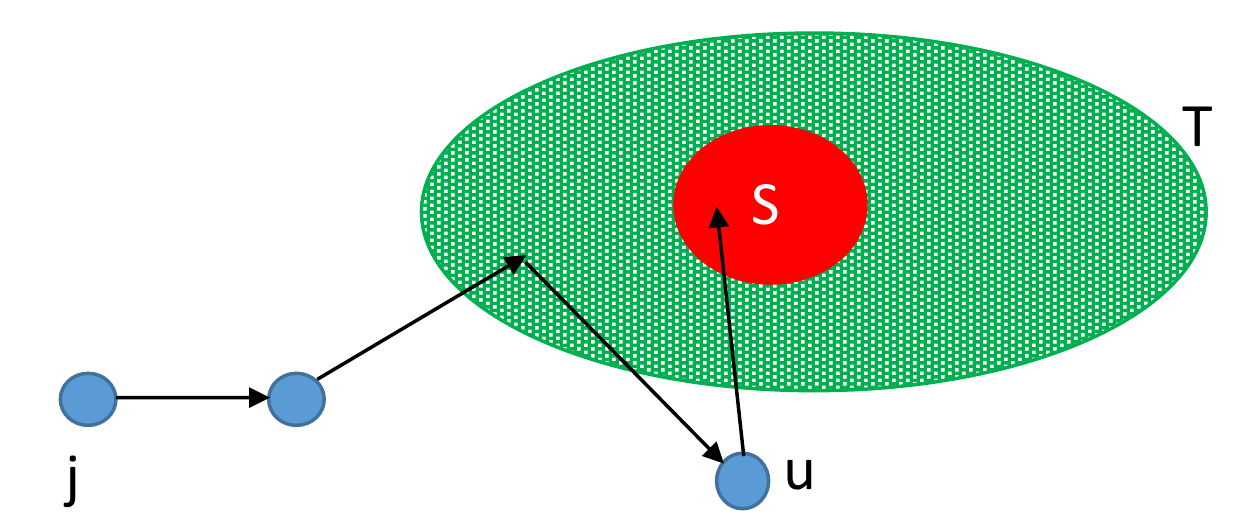} \\
\mbox{(a)} & \mbox{(b)} & \mbox{(c)}
\end{array}$
\caption{Illustrations of the supermodularity of commute time. (a) If the walk reaches $j$ before $T$, then it automatically reaches $j$ before $S$, implying that $\tau_{j}(T) \subseteq \tau_{j}(S)$. (b) When a walk starting at $j$ reaches $S$ before $u$, the time for the walk to start at $j$, reach $S$, and then reach $u$ ($U_{jSu}$) is equal to the corresponding time $U_{jTu}$ for set $T$. (c) The case where $U_{jSu} > U_{jTu}$. In this case, the walk reaches $T \setminus S$ and then reaches node $u$ before reaching any node in $S$.}
\label{fig:commute_supermodular}
\end{figure}

To see that $Pr(\tau_{j}(S)) \geq Pr(\tau_{j}(T))$, note that if the walk reaches $j$ before $T$, then it automatically reaches $j$ before $S$ since $S \subseteq T$ (Figure \ref{fig:commute_supermodular}(a)). Turning to the inequality $U_{jSu} \geq U_{jTu}$, this property can be shown by considering the cases in Figure \ref{fig:commute_supermodular}(b)--\ref{fig:commute_supermodular}(c). 

First, if the walk starting at $j$ reaches $S$ before reaching $u$, then $U_{jSu} = U_{jTu}$ (Figure \ref{fig:commute_supermodular}(b)). On the other hand, if the walk reaches $T \setminus S$ and then reaches $u$ before $S$, then $U_{jSu} > U_{jTu}$ (Figure \ref{fig:commute_supermodular}(b)). Hence $U_{jSu} \geq U_{jTu}$ in all cases, completing the proof of supermodularity.

\begin{figure}[!ht]
\centering
$\begin{array}{ccc}
\includegraphics[width=2in]{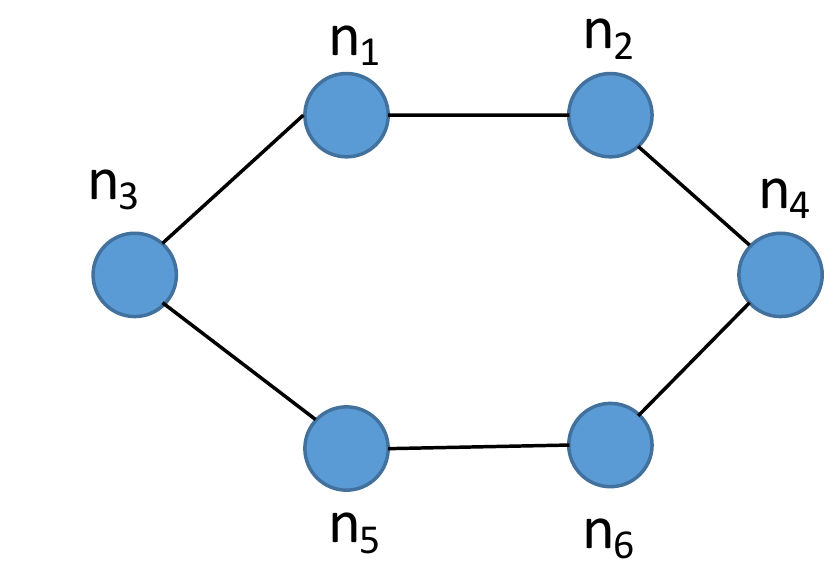} &
\includegraphics[width=2in]{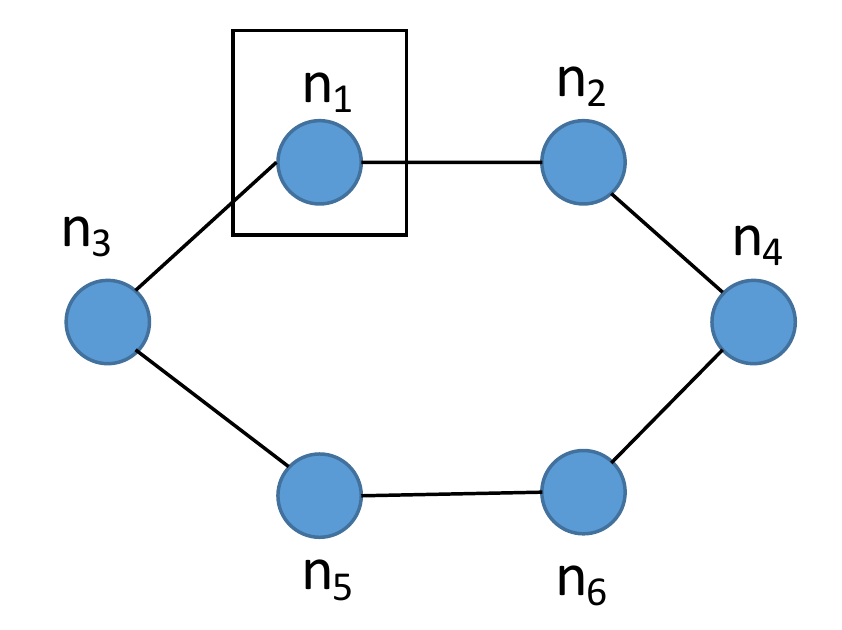} &
\includegraphics[width=2in]{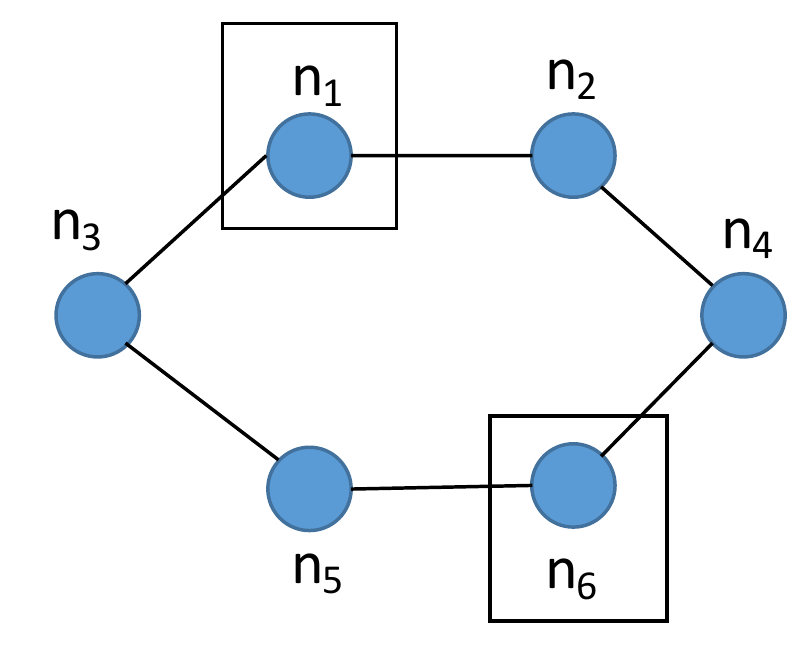} \\
\mbox{(a)} & \mbox{(b)} & \mbox{(c)} 
\end{array}$
\caption{Example of input selection for robustness to noise. (a) Initial network topology. (b) Input selected at first iteration of greedy algorithm. (c) Input selected at second iteration.}
\label{fig:noise_example}
\end{figure}

The supermodularity property implies that an efficient greedy algorithm is sufficient to minimize the robustness to noise up to a provable optimality bound of $(1-1/e)$. Consider the ring topology shown in Figure \ref{fig:noise_example}(a), with all links having equal edge weight of $1$. Suppose that $k=2$ input nodes are selected according to the submodular (greedy) algorithm. By symmetry, all input nodes  provide the same error due to noise at the first iteration of the algorithm. Without loss of generality, suppose that $n_{1}$ is selected (Figure \ref{fig:noise_example}(b)). At the second iteration, the node $v$ that minimizes $R(\{n_{1},v\})$  is equal to $n_{6}$ (Figure \ref{fig:noise_example}(c)). The greedy algorithm in this case provides the optimal solution to the input selection problem \cite{patterson2015efficient}. 

\subsection{Submodular Approach for Kalman Filtering}
\label{subsec:Kalman}
 This section presents submodular techniques for ensuring that estimation processes are robust to error, with a focus on the classical Kalman filter. Consider a discrete-time system with state dynamics
\begin{eqnarray*}
x_{k+1} &=& A_{k}x_{k} + w_{k} \\
y_{k} &=& C_{k}x_{k} + v_{k}
\end{eqnarray*}
with initial condition $x_{0}$, where $w_{k}$ and $v_{k}$ are process and measurement noise, respectively. The Kalman filter develops a minimum mean-square estimate of $x_{k}$ over an observation interval $\{0,\ldots,k\}$. Suppose that the estimation errors at different time steps are independent and identically distributed, and that the covariance matrix of $v_{k}$ is equal to $\sigma^{2}I$ for some $\sigma \in \mathbb{R}$. The estimator produces an estimate of $z_{k} \triangleq (x_{0}^{T} \ w_{k}^{T})^{T}$ given by $$\hat{z}_{k-1} \triangleq \mathbf{E}(z_{k}) + \mathbf{C}(z_{k-1})O_{k}^{T}(O_{k}\mathbf{C}(z_{k-1})O_{k}^{T} + \sigma^{2}I)^{-1}(\overline{y}_{k} - O_{k}\mathbf{E}(z_{k-1}) - \mathbf{E}(\overline{v}_{k})),$$ where $$O_{k} = (L_{0}^{T}C_{0}^{T} \ L_{1}^{T}C_{1}^{T} \ \cdots \ L_{k}^{T}C_{k}^{T})^{T}, \quad L_{i} = (A_{i-1} \cdots A_{0}, A_{i-1} \cdots A_{1}, \ldots A_{i-1}, I, 0),$$ $\mathbf{C}$ denotes the covariance matrix, $\overline{v}_{k} = (v_{0} \ v_{1} \ \cdots v_{k})$, and $\overline{y}_{k} = O_{k}z_{k-1} + \overline{v}_{k}$. 

The performance of the Kalman filter has been well-studied~\cite{grewal2011kalman}. The covariance matrix of the estimate, denoted $\Sigma_{z_{k-1}}$, is known to be given by 
\begin{equation}
\label{eq:KF_cov}
\Sigma_{z_{k-1}} = \mathbf{C}(z_{k-1}) - \mathbf{C}(z_{k-1})O_{k}^{T}(O_{k}\mathbf{C}(z_{k-1})O_{k}^{T} + \sigma^{2}I)^{-1}O_{k}\mathbf{C}(z_{k-1}),
\end{equation}
for a minimum mean-square error of $\mathbf{tr}(\Sigma_{z_{k-1}})$.

An additional performance metric is the volume of the $\eta$-confidence ellipsoid, equal to the minimum volume ellipsoid that contains $z_{k-1} - \hat{z}_{k-1}$ with probability $\eta$. Hence, a smaller-volume ellipsoid implies a reduced error of the filter. The volume of the ellipsoid is given by $$\frac{(\eta \pi)^{n(k+1)/2}}{\Gamma(n(k+1)/2 + 1)}\det{\left(\Sigma_{z_{k-1}}^{1/2}\right)}.$$ Equivalent to minimizing this volume is minimizing the logarithm of the volume, equal to $\beta + 1/2 \log{\det{(\Sigma_{z_{k-1}})}}$, where $\beta$ depends only on $n(k+1)$ and $\eta$. 

The estimation error can be varied by choosing the set of state variables that are observed by the matrix $C$~\cite{tzoumas2015sensor}. Under this model, the matrix $C$ has one nonzero entry per row, corresponding to an input node that sends a measurement. The input selection problem is then formulated as $\min{\{h(S) \triangleq \log{\det{\Sigma_{z_{k-1}}}} : |S| \leq r\}},$ where $r$ is the number of inputs and $S$ is the set $\{j : C_{ij} = 1 \mbox{ for some $i$ }\}$. The following result leads to a submodular approach to approximately solving this input selection problem.


\begin{theorem}[\cite{tzoumas2015sensor}]
\label{theorem:log_det_submod}
The function $h(S) = \log{\det{\Sigma_{z_{k-1}}}}$ is supermodular. 
\end{theorem}

Supermodularity of the metric $h(S)$ implies that a set of input nodes can be selected to minimize the error of Kalman filtering using a supermodular optimization approach with provable optimality bounds. 


\section{Submodularity for Performance: Smooth Convergence}
\label{sec:convergence}
Even if a networked system is guaranteed to converge asymptotically to a desired state, the deviations of the intermediate states from the steady-state value may lead to undesirable system performance. In the formation maneuvering example of the previous section, intermediate state deviations correspond to position errors of the nodes. Control of power systems requires the system to not only reach a stable operating point, but also reach that point in a timely fashion.

The convergence of a network where the input node states are constant and the non-input nodes have dynamics 
\begin{equation}
\label{eq:weighted_consensus}
\dot{x}_{i}(t) = \sum_{j \in N_{in}(i)}{W_{ij}(x_{i}(t)-x_{j}(t))},
\end{equation}
which generalize the consensus dynamics of (\ref{eq:consensus}) to directed graphs, is analyzed as follows. Asymptotically, the non-input node states  achieve \emph{containment}~\cite{ji2008containment, cao2012distributed}, defined as all node states lying in the convex hull of the input node states (Figure \ref{fig:containment}). Containment is a natural property in coordinated motion and coverage problems, where ensuring that all network nodes remain within a desired region is a necessary requirement. Containment may also be needed to ensure that the network remains connected in steady-state. In the special case where all non-input nodes have the same state, containment is equivalent to consensus.

Motivated by the containment property, the convergence error of a networked system at time $t$ is defined by $f_{t}(S) = \mbox{dist}(\mathbf{x}(t), \overline{A})^{p},$ where $S$ denotes the input set and $\overline{A}$ denotes the convex hull of the input node positions, and $p \geq 1$ is a parameter of the error metric. Here, $\mbox{dist}(\cdot,\overline{A})$ is the distance from a point to set $\overline{A}$ with respect to any $l_{p}$-norm.

As in the case of error due to noise, the submodular design for smooth convergence is based on establishing a connection to a random walk on the network graph. The key insight is that the dynamics (\ref{eq:weighted_consensus}) define a diffusion process on the graph, in which differences between initial state values are diffused among neighboring nodes until all node states have the same value. The relationship between diffusion processes and random walks has been well-studied. For instance, a classical result states that solutions to the heat equation are equal to the expected value of a Brownian motion~\cite{karatzas2012brownian}.

The following is a discrete analog of the Brownian motion-heat equation relationship, in which a random walk on a graph is analogous to a Brownian motion in $\mathbb{R}^{n}$, that will be used to develop submodular techniques for smooth convergence.  As described in the section ``Submodularity for Performance: Robustness to Noise,'' the weighted averaging dynamics can be written in the form $\dot{\mathbf{x}}(t) = -L\mathbf{x}(t)$, where $L$ denotes the graph Laplacian matrix, so that $\mathbf{x}(t) = e^{-Lt}\mathbf{x}(0)$. It can be shown that $e^{-Lt}$ is a stochastic matrix for any $t > 0$, and hence is the transition matrix of a random walk~\cite{mesbahi2010graph}. Furthermore, for any node $i \in S$, the $i$-th diagonal entry of $e^{-Lt}$ is equal to $1$, implying that the input nodes are \emph{absorbing states} of the walk (see ``Random Walks on Graphs'' for details).

\begin{figure}[!ht]
\centering
\includegraphics[width=3in]{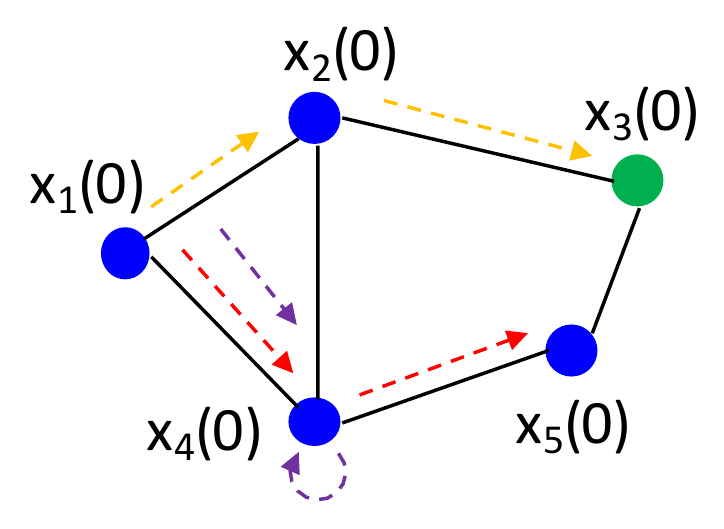}
\caption{Connection between node dynamics and a random walk on the graph. The value of $x_{i}(t)$ is equal to the expected initial state value of the node reached by the walk. As an example, if $x_{3}(0) = 1$, $x_{5}(0) = 2$, and $x_{4}(0) = 0.6$, then the  value of $x_{1}(t)$ based on the three sample paths shown would be $(1+2+0.6)/3 = 1.2$.}
\label{fig:RW_diffusion}
\end{figure}

Suppose that $x_{i}(0)$ denotes the initial state of node $i$, and let $X_{i}[k]$ be a random walk with transition matrix $P=e^{-L\delta}$ and $X[0] = i$. For any $t = k\delta$, $x_{i}(t)$ is equal to $e_{i}^{T}P^{k}\mathbf{x}(0)$, which is exactly equal to $\mathbf{E}(x_{X[k]}(0))$~\cite{clark2014minimizing}. This is equivalent to the expected value of $x(0)$ at the node reached at step $k$ of the walk (Figure \ref{fig:RW_diffusion}).

A consequence of this relationship is that, intuitively, the state of node $i$ ``converges'' when the random walk reaches the input set $S$. This intuition can be formalized through the inequality
\begin{equation}
\label{eq:conv_error_bound}
f_{t}(S) \leq K \left(\sum_{i \in V \setminus S}{\left[\sum_{j \in V \setminus S}{g_{ij}(S)^{p}} + h_{i}(S)^{p}\right]}\right)^{1/p},
\end{equation}
where
\begin{equation*}
g_{ij}(S) = Pr(X(\tau)=j | X(0)=i), \quad  h_{i}(S) = Pr(X(\tau) \notin S | X(0) = i)
\end{equation*}
and $K$ is an upper bound determined by the norm of $\mathbf{x}(0)$~\cite{clark2014minimizing}. The function $g_{ij}(S)$ is the probability that a walk starting at node $i$ reaches node $j$ after $\tau$ steps, while $h_{i}(S)$ is the probability that a walk starting at node $i$ does not reach the input set at step $\tau$. The choice of input set impacts the values of $g_{ij}(S)$ and $h_{i}(S)$ because each input node is an absorbing state of the walk.

\begin{figure}[!ht]
\centering
$\begin{array}{ccc}
\includegraphics[width=2in]{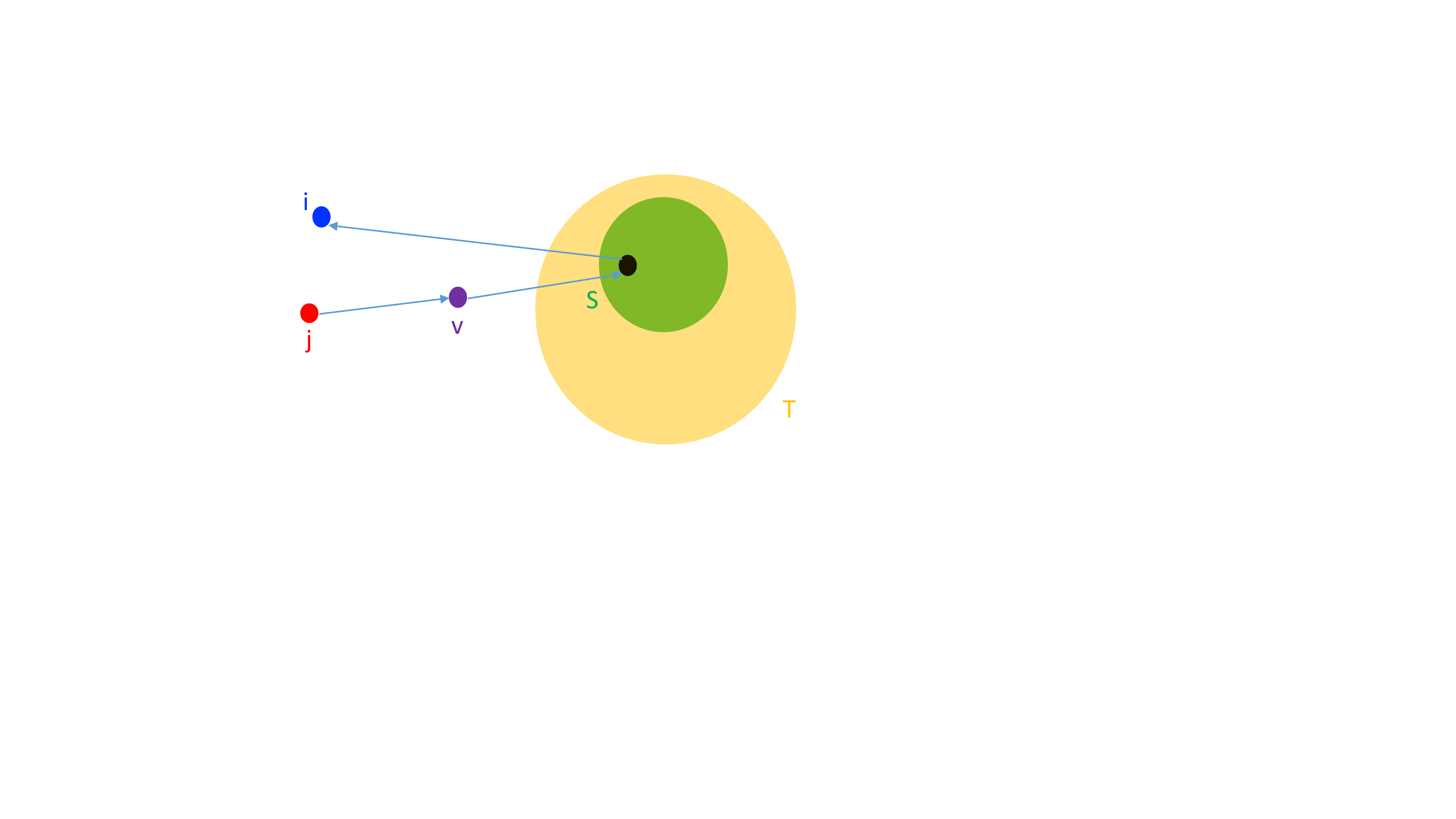} &
\includegraphics[width=2in]{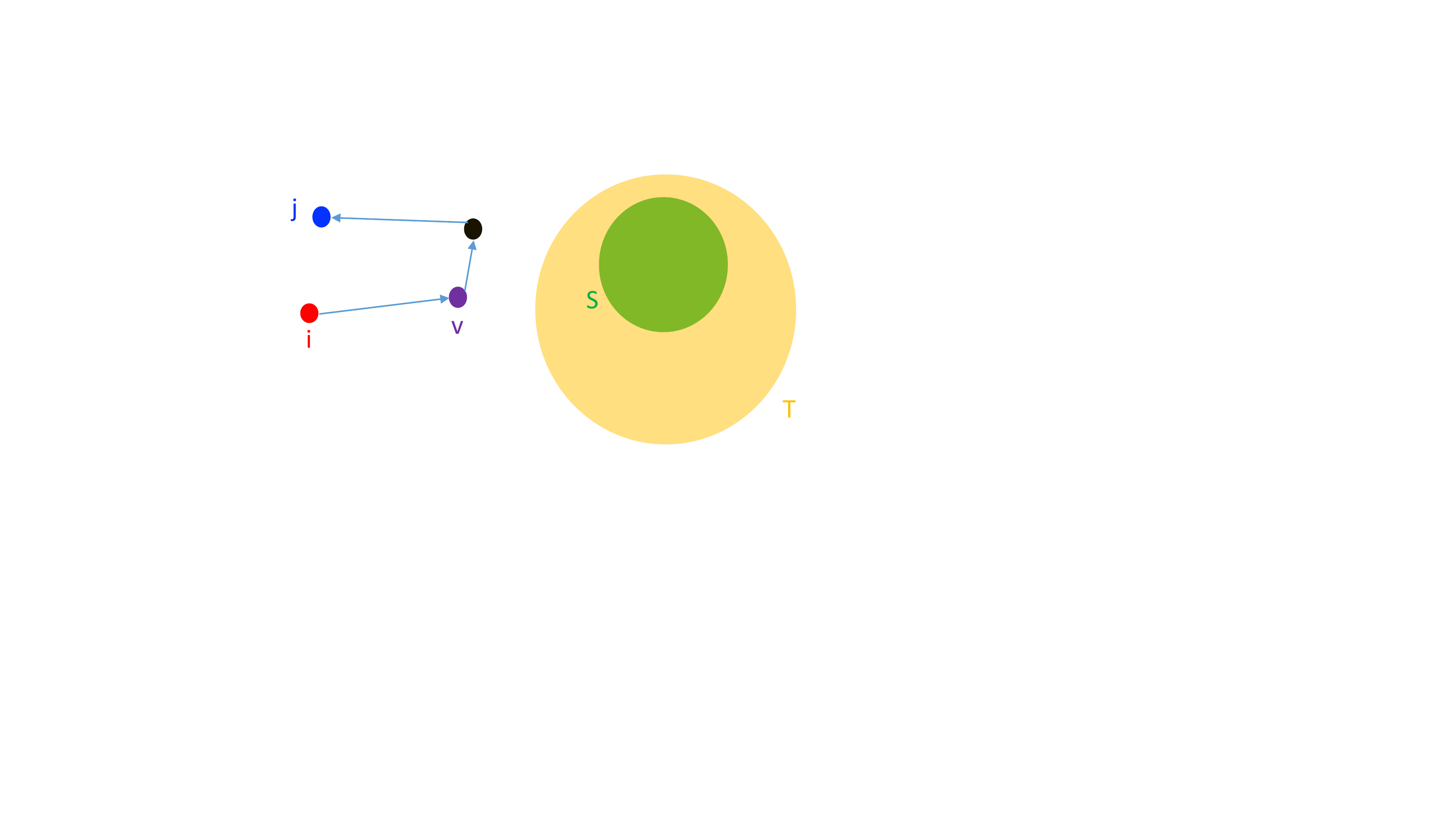} &
\includegraphics[width=2in]{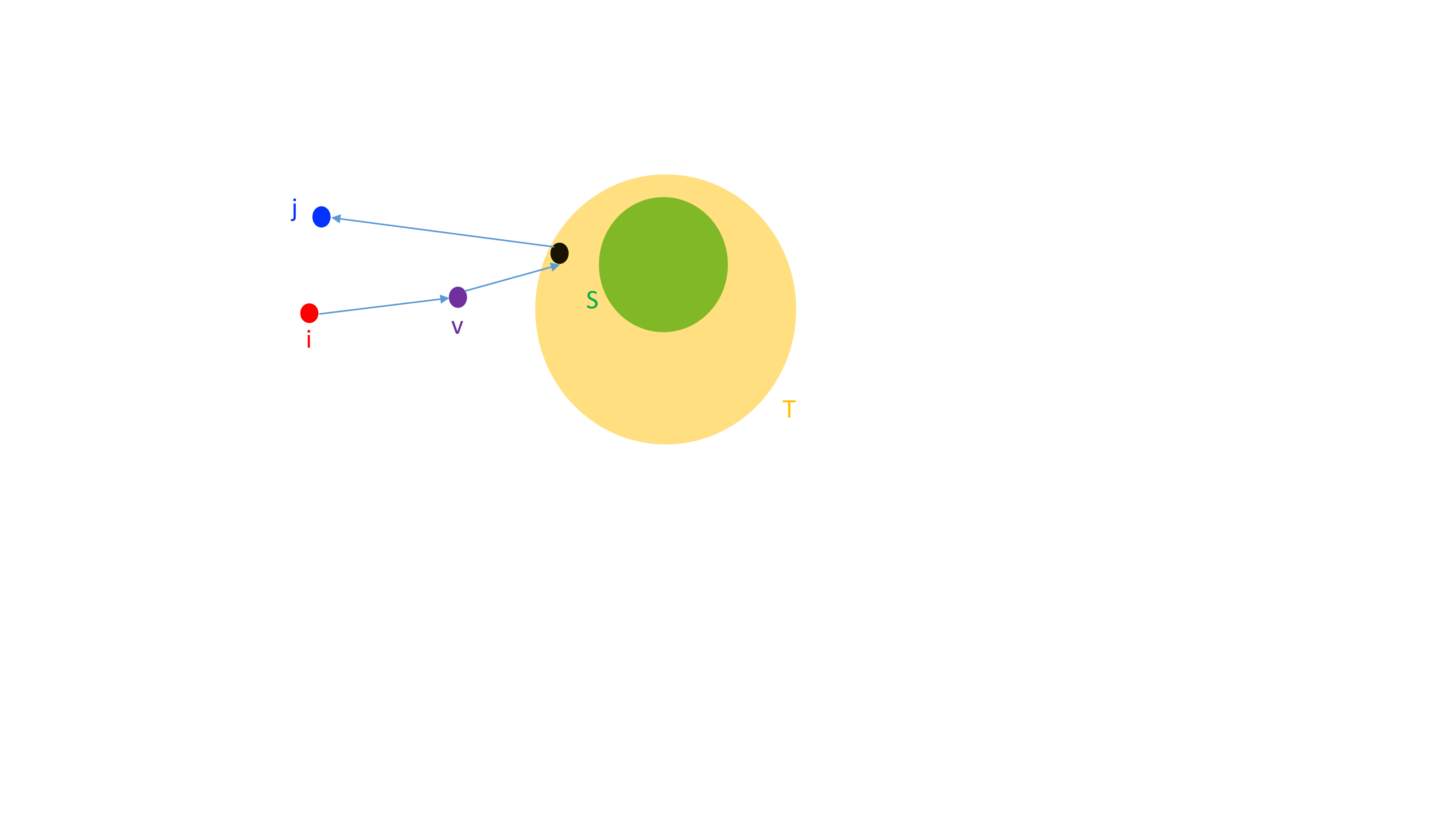} \\
\mbox{(a)} & \mbox{(b)} & \mbox{(c)} 
\end{array}$
\caption{Three cases to illustrate supermodularity of the convergence error as a function of the input set. (a) The walk reaches the set $S$, and hence both sides of (\ref{eq:convergence_supermod}) are zero. (b) The walk reaches $v$ but not $S$, implying that both sides of (\ref{eq:convergence_supermod}) are equal. (c) The walk reaches $v$ and $T \setminus S$ but not $S$, implying that (\ref{eq:convergence_supermod}) holds with strict inequality.}
\label{fig:convergence_supermodular}
\end{figure}

In order to develop a submodular approach to selecting input nodes for smooth convergence, it suffices to prove supermodularity of $g_{ij}(S)$ and $h_{i}(S)$. The supermodular structure arises because the set of input nodes $S$ represents a set of absorbing states of the walk. If a walk reaches a node in $S$ at step $\tau^{\prime} < \tau$, then the walk remains at that node and does not reach $j \in V \setminus S$ at step $\tau$. 
Thus the increment $g_{ij}(S)-g_{ij}(S \cup \{v\})$ is equal to the probability that a walk starting at $i$ reaches $v$ and $j$ by time $\tau$, but does not reach the set $S$. The inequality 
\begin{equation}
\label{eq:convergence_supermod}
g_{ij}(S) - g_{ij}(S \cup \{v\}) \geq g_{ij}(T) - g_{ij}(T \cup \{v\})
\end{equation}
 can be shown by considering separate cases of $S$ and $T$, illustrated in Figure \ref{fig:convergence_supermodular}. 

If the walk reaches $S$ (Figure \ref{fig:convergence_supermodular}(a)), then both sides of (\ref{eq:convergence_supermod}) are automatically zero, since the probability that the walk continues to node $j$ after reaching $S$ is zero. Similarly, if the walk does not reach $T$, then the input set does not impact the walk and hence both sides of (\ref{eq:convergence_supermod}) are equal (Figure\ref{fig:convergence_supermodular}(b)).  In the case shown in Figure \ref{fig:convergence_supermodular}(c),  the walk reaches $T$ and $v$ but not $S$ within $\tau$ steps. Hence the left-hand side of (\ref{eq:convergence_supermod}) is positive, while the right-hand side is zero, implying that (\ref{eq:convergence_supermod}) holds with strict inequality. 

Combining these arguments and the composition rules for supermodular functions yields the following main result.
\begin{theorem}[\cite{clark2014minimizing}]
\label{theorem:convergence_supermod}
The convergence error bound $$\hat{f}_{t}(S) = \sum_{i \in V \setminus S}{\left[\sum_{j \in V \setminus S}{g_{ij}(S)^{p}} + h_{i}(S)^{p}\right]}$$ is supermodular as a function of the input set $S$.
\end{theorem}

The supermodularity of the convergence error implies that efficient algorithms can be developed for selecting input nodes to minimize the deviation of the intermediate node states. Moreover, Theorem \ref{theorem:convergence_supermod} implies that generalizations of the convergence error, such as the integral $\int_{0}^{\infty}{\hat{f}_{t}(S) \ dt}$, are supermodular functions of the input set~\cite{clark2014minimizing}.

\section{Input Selection in Dynamic Networks}
\label{subsec:dynamic}
The discussion so far has implicitly assumed that the network topology $G=(V,E)$ is fixed over time. The topologies of networked systems, however, evolve over time. The model for the network topology dynamics naturally depends on the cause of these variations. In many cases, however, the submodular framework extends naturally to dynamic topologies.

One source of topology changes is random failures of nodes or links in an otherwise static topology. Node failures may occur due to hardware failures, or participants dropping out of a social network, while link failures often occur due to communication over lossy wireless channels. In both cases, for a given performance metric $f(S)$, the effect of the topology can be quantified as the expected value of the metric, $\overline{f}(S) = \mathbf{E}_{\pi}(f(S))$, where $\pi$ denotes the probability distribution on the network topology due to node and link failures. The following result enables extending the submodular design approach to network topologies with random failures.

\begin{lemma}[\cite{kempe2003maximizing}]
If $f(S)$ is submodular (respectively, supermodular), then the function $\overline{f}(S) = \mathbf{E}_{\pi}(f(S))$ is a submodular (respectively, supermodular) function of $S$.
\end{lemma}
 The proof can be seen from the fact that $\overline{f}(S)$ is a nonnegative weighted sum of submodular (or supermodular) functions. While submodularity of $f(S)$ enables provable guarantees for simple greedy algorithms, the complexity of evaluating the function $\overline{f}(S)$ is worst-case exponential, since all possible network topologies may have nonzero probability. Monte Carlo methods or approximations to specific cost functions may be used in this case \cite{clark2014supermodular}.
 
The network topology may also undergo changes caused by switching between predefined topologies. Switching topologies are common in formation maneuvers, where different topologies are used to change the coverage area of the formation and avoid obstacles \cite{olfati2004consensus, mesbahi2001formation}. 

The effect of the switching can be modeled as a set of topologies $\{G_{1},\ldots,G_{M}\}$. Two relevant metrics are the \emph{average} and \emph{worst-case} performance. The average case performance, given as $f_{avg}(S) = \frac{1}{M}\sum_{i=1}^{M}{f(S|G_{i})}$, inherits the submodular structure of the objective function $f(S)$, similar to the case of random failures. 

The worst-case performance is formulated as $f_{worst}(S) = \max{\{f(S|G_{i}) : i=1,\ldots,M\}}$, and unlike $f_{avg}$ is not supermodular as a function of $S$. The problem of selecting a minimum-size input set to achieve a desired bound on the worst-case performance can, however, be approximated with provable optimality guarantees by using the equivalent formulation~\cite{krause2007selecting, clark2014supermodular,clark2014minimizing}
\begin{equation}
\label{eq:worst_case}
\begin{array}{ll}
\mbox{minimize} & |S| \\
\mbox{s.t.} & \max_{i=1,\ldots,M}{f(S|G_{i})} \leq \alpha
\end{array}
\Leftrightarrow
\begin{array}{ll}
\mbox{minimize} & |S| \\
\mbox{s.t.} & \frac{1}{M}\sum_{i=1}^{M}{\max{\{f(S|G_{i}), \alpha\}}} \leq \alpha
\end{array}
\end{equation}
The function $\max{\{f(S),c\}}$ is supermodular whenever $f(S)$ is a decreasing supermodular function and $c$ is a real constant, and hence the equivalent problem formulation defines a supermodular optimization problem.

\section{Submodularity and Controllability}
\label{sec:controllability}
A networked system is controllable if it is possible to drive the node states $\mathbf{x}(t)$ from any initial values $\mathbf{x}(0)$ to any desired final values $\mathbf{x}(T)$ in a finite time $T$. The problem of selecting input nodes to guarantee controllability has received significant attention, including results on controllability of consensus networks \cite{tanner2004controllability,rahmani2009controllability, goldin2013weight}, networks with known parameters \cite{summers2016controllability}, and topologies with known and unknown parameters (structured systems) \cite{liu2011controllability,clark2012controllability, olshevsky2014minimal, pequito2015minimum}. This section presents submodular optimization techniques for selecting input nodes for controllability. 

\subsection{Controllability of Networked Systems}
\label{subsec:controllability_known}
Conditions for controllability of linear systems have been studied since the 1960s, when Kalman's controllability criteria were presented. It was shown that a linear system 
\begin{equation}
\label{eq:linear_system}
\dot{\mathbf{x}}(t) = A\mathbf{x}(t) + B\mathbf{u}(t)
\end{equation}
with $\mathbf{x}(t) \in \mathbb{R}^{n}$ 
is controllable if and only if the controllability matrix $\mathcal{C} = (B \ AB \ A^{2}B \ \cdots \ A^{n-1}B)$ 
 has full rank. 
 
 In a networked system, the input nodes affect the matrix $\mathcal{C}$ by determining the columns of the $B$ matrix. If the dynamics of the system in the absence of any inputs are given by $\dot{\mathbf{x}}(t) = W\mathbf{x}(t)$, then selecting an input set $S$ results in $A = (W_{ij} : i, j \in V \setminus S)$ and $B = (W_{ij} : i \in V \setminus S, j \in S)$. Based on this insight, the problem of selecting a minimum-size set of input nodes to ensure controllability can be formulated as 
 \begin{equation}
 \label{eq:controllability_select}
 \begin{array}{ll}
 \mbox{minimize} & |S| \\
 \mbox{s.t.} & r(S) = n
 \end{array}
 \end{equation}
where  $r(S) = \mbox{rank}(\mathcal{C}(S))$ and $\mathcal{C}(S)$ is the controllability matrix when the input set is $S$.
 The following result leads to submodular approaches to selecting input nodes for controllability.
 \begin{lemma}[\cite{summers2016controllability}]
 \label{lemma:controllability_matrix}
 The function $r(S)$ is a monotone submodular function of $S$.
 \end{lemma}


By Lemma \ref{lemma:controllability_matrix}, a simple greedy algorithm suffices to select a set of input nodes with provable bounds on the cardinality of the set. Indeed, by applying the bounds on the greedy algorithm for submodular cover, it follows that the set $S$ selected by the greedy algorithm satisfies $$\frac{|S|}{|S^{\ast}|} \leq 1 + \log{n},$$ where $S^{\ast}$ denotes the optimal solution. Furthermore, it can be shown that this is the best bound that can be achieved unless the P=NP conjecture from complexity theory holds~\cite{olshevsky2014minimal}.


In addition to the controllability of the system, the controllability matrices provide insight into the amount of energy that must be exerted to control a networked system. One such controllability matrix is the controllability Gramian, defined as the positive semidefinite solution $W_{S}$ to $$AW_{S} + W_{S}A^{T} + B_{S}B_{S}^{T} = 0.$$ The $H_{2}$ norm of the system is a weighted trace of the controllability Gramian, so that $||H||_{2}^{2} = \mathbf{tr}(XW_{S}X^{T})$ for some matrix $X$. The trace, in turn, is a modular function of the set of input nodes~\cite{summers2016controllability}.

An additional energy-related metric is the trace of the \emph{inverse} of the controllability Gramian. This metric is proportional to the energy needed on average to steer the system from the initial operating point to the final, desired state. The trace of the inverse of the controllability Gramian is a monotone decreasing and supermodular function of the input set $S$~\cite{summers2016controllability}.

\subsection{Structural Controllability}
\label{subsec:SC}
In the preceding analysis, it was assumed that all parameters, such as interaction weights, between nodes are known a priori. In many systems of interest, such as biological networks, the parameters cannot be observed directly, or are estimated with errors or uncertainties. In such systems, controllability can still be analyzed by considering the \emph{structural rank} of the system.

\begin{definition}[\cite{lin1974structural}]
\label{def:structural_rank}
The structural rank of a system is defined as the maximum rank of the controllability matrix over all values of the weights $W_{ij}$ in (\ref{eq:dynamics}). A system satisfies \emph{structural controllability} if the structural rank of the controllability matrix is equal to the number of non-input nodes.
\end{definition}

Choosing the structural rank as the maximum achievable rank may seem optimistic. It can, however, be shown that \emph{any} set of parameters $W_{ij}$ achieve the structural rank, except when the weights  are chosen from a set that has Lebesgue measure zero~\cite{lin1974structural}.  Stricter conditions have also been formulated; in \cite{chapman2015strong}, conditions for strong structural controllability, which implies controllability for any nonzero values of the free parameters, are presented. Controllability conditions for linear descriptor systems, which have a combination of free and fixed parameters, are discussed in \cite{murota2012systems}. Furthermore, other structural conditions have been proposed, including disturbance rejection properties~\cite{willems1981disturbance}, which can be relaxed to matroid constraints~\cite{clark2015input}.  In what follows, however, the analysis focuses on structural controllability as in Definition \ref{def:structural_rank}.

\begin{figure}[!ht]
\centering
\includegraphics[width=2.5in]{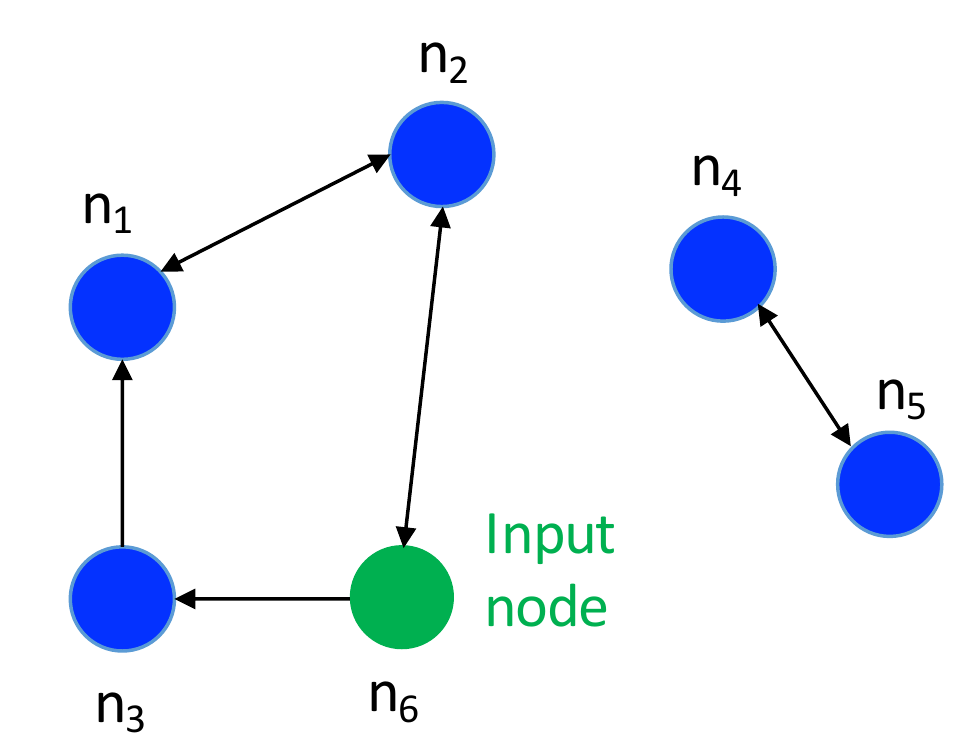}
\caption{The accessibility condition. The nodes $n_{4}$ and $n_{5}$ are not accessible from the input node $n_{6}$, and hence the graph is not controllable.}
\label{fig:accessibility}
\end{figure}

An advantage of structural controllability is that it can be characterized using properties of the network graph, enabling a common set of techniques to be used to analyze systems in different application domains. To motivate one necessary condition for structural controllability, consider the graph shown in Figure \ref{fig:accessibility}.

By inspection, the $A$ and $B$ matrices from Eq. (\ref{eq:linear_system}) arising from this graph (which has input node $n_{6}$) are of the form 
\begin{displaymath}
A = \left(
\begin{array}{ccccc}
0 & * & * & 0 & 0 \\
* & 0 & 0 & 0 & 0 \\
0 & 0 & 0 & 0 & 0 \\
0 & 0 & 0 & 0 & * \\
0 & 0 & 0 & * & 0
\end{array}
\right), \quad B = \left(
\begin{array}{c}
0 \\
* \\
* \\
0 \\
0 
\end{array}
\right)
\end{displaymath}
and hence the states of nodes $n_{4}$ and $n_{5}$ are not controllable. From the graph-theoretic viewpoint, the nodes $n_{4}$ and $n_{5}$ are not connected to the input node, and hence cannot be controlled from that node. This motivates the accessibility property, defined as follows.


\begin{definition}
\label{def:accessibility}
A node in a networked system is \emph{accessible} if there is a path from an input node to that node. The system satisfies accessibility if all nodes are accessible.
\end{definition}

\begin{figure}[!ht]
\centering
\includegraphics[width=2.5in]{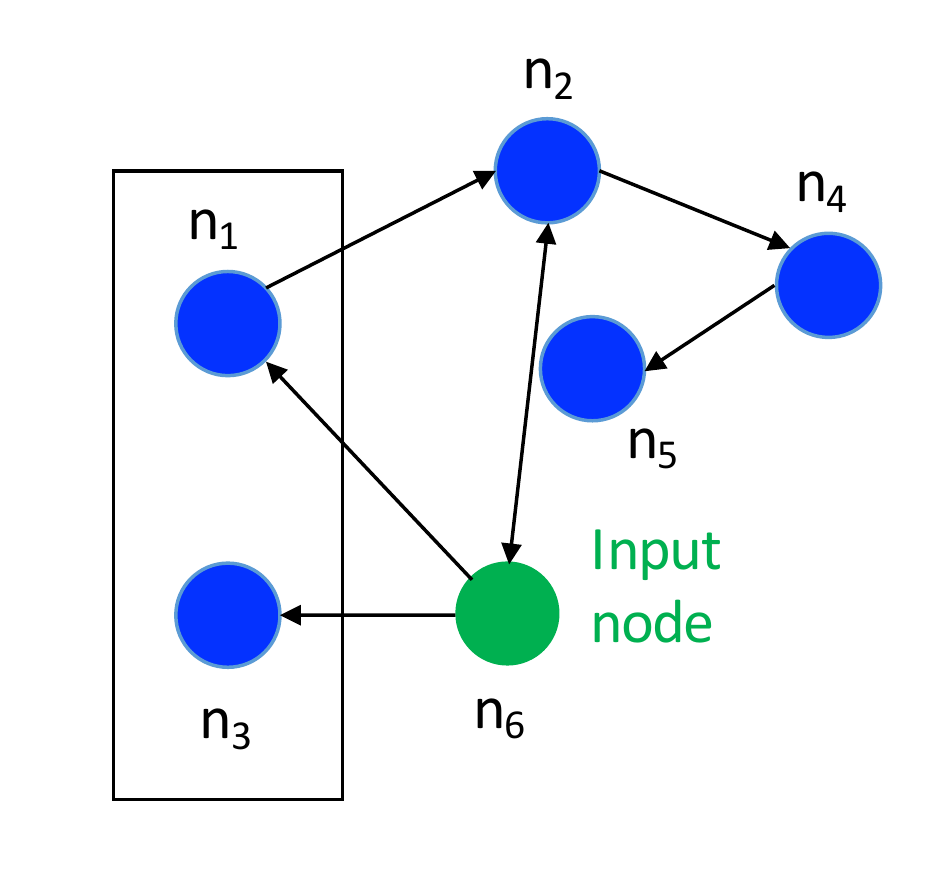}
\caption{Network that contains a dilation. The system is not controllable because the two nodes $\{n_{1},n_{3}\}$ have only one neighbor, namely the input node $n_{6}$. In this case, $N(A) = \{n_{6}\}$ and $|A| > |N(A)|$.}
\label{fig:DF}
\end{figure}

Accessibility is a necessary condition for controllability. For the next controllability condition, consider the example of Figure \ref{fig:DF}.

In the graph of Figure \ref{fig:DF}, the nodes $n_{1}$ and $n_{3}$ both have exactly one neighbor, the input node $n_{6}$. Hence the state dynamics of node $n_{1}$ and $n_{3}$ satisfy $\dot{x}_{1}(t) = \alpha \dot{x}_{3}(t)$ for some constant $\alpha \in \mathbb{R}$, and the states $x_{1}(t)$ and $x_{3}(t)$ always lie in an affine subspace of $\mathbb{R}^{2}$ that is determined by the initial state and $\alpha$. It is therefore impossible to drive $x_{1}(t)$ and $x_{3}(t)$ to any arbitrary values, implying that controllability does not hold.
 We generalize this property to the notion of dilation-freeness.


\begin{definition}
\label{def:DF}
A network is \emph{dilation-free} if, for any set of nodes $A \subseteq V$, $|N(A)| \geq |A|$, where $N(A) = \cup_{i \in A}{N_{in}(i)}$ (in words, the number of neighbors of $A$ is at least as large as the set $A$).
\end{definition}

Dilation-freeness can be interpreted by the following intuition. For a set of $m$ nodes, in order for those nodes to be driven to any arbitrary state, at least $m$ degrees of freedom would be needed. Otherwise, the inputs received by any two nodes would satisfy a linear relationship, implying that the vector of node states $\mathbf{x}(t)$ would lie in a subspace of $\mathbb{R}^{n}$. In Figure \ref{fig:DF}, the set $A = \{n_{1}, n_{3}\}$ and $N(A) = \{n_{6}\}$, hence violating dilation-freeness.

Both accessibility and dilation-freeness are necessary conditions for structural controllability. It can also be shown that the converse is true.

\begin{theorem}[\cite{lin1974structural}]
\label{theorem:SC}
If a networked system satisfies accessibility and dilation-freeness, then the system is structurally controllable.
\end{theorem}

The dilation-free property also has a connection to matching theory (see Graph Matchings sidebar), which can be understood using the Hall Marriage Theorem.

\begin{theorem}[Hall Marriage Theorem~\cite{lovasz2009matching}]
\label{theorem:Hall}
For any bipartite graph, there exists a perfect matching if and only if each set $A \subseteq V$ satisfies $|N(A)| \geq |A|$. 
\end{theorem}

From Theorem \ref{theorem:Hall}, the dilation-free property is equivalent to the existence of a perfect matching in the bipartite graph $G = (U,Z,E)$, where $Z = V \setminus S$ (the set of non-input nodes), $U$ is the set of all nodes in the network, and the edge $(u_{i},z_{j})$ exists if $(i,j) \in E$. 

\begin{figure}
 \centering
 \includegraphics[width=6in]{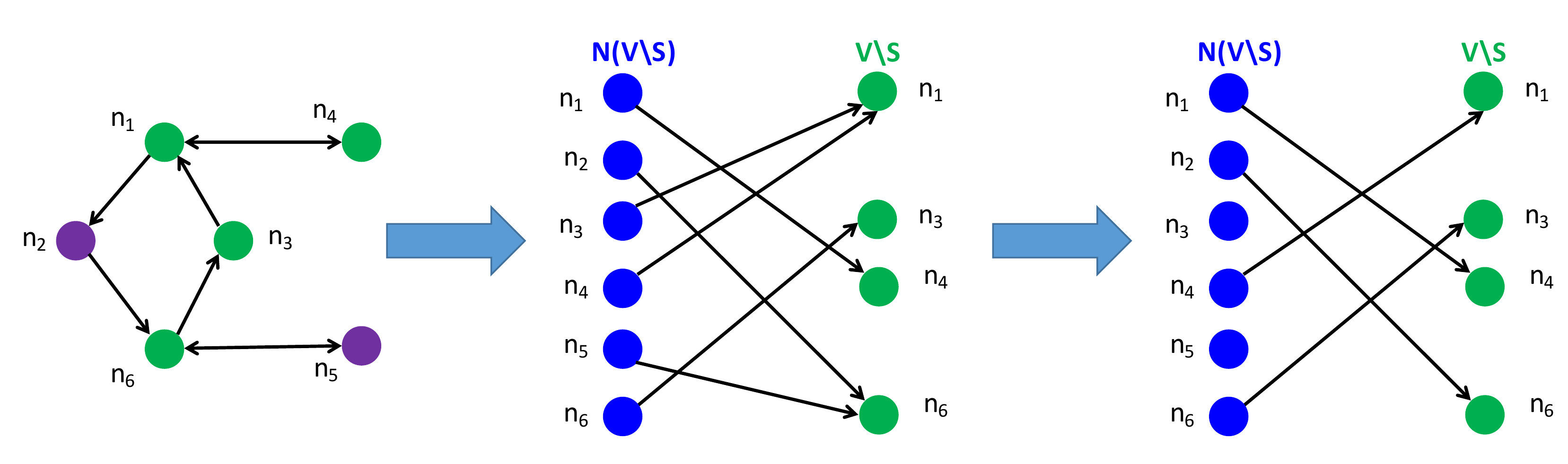}
 \caption{Mapping controllability to a matching constraint. The first step is to map the network graph to a bipartite representation, and then construct a maximal matching. Since there is a matching from $N(V \setminus S)$ into $(V \setminus S)$ in which all nodes in $V \setminus S$ are matched, the graph is controllable from input set $S=\{n_{2},n_{5}\}$.}
 \label{fig:controllability_matching}
 \end{figure}

 In the case where the graph is strongly connected, the minimum-size set of input nodes to guarantee structural controllability can be chosen based on a graph-matching algorithm. Under the algorithm, a maximum matching on the graph is computed using a technique such as the Hungarian algorithm~ \cite{lovasz2009matching}. All nodes that are left unmatched under the maximum matching are then chosen as inputs. The connection between graph matchings and controllability is illustrated in the example of Figure \ref{fig:controllability_matching}.
 
The matching condition on controllability can be expressed as a matroid constraint through the following analysis. The mapping to matroids is a step towards developing joint input selection algorithms for performance and controllability.


Consider the problem of selecting a feasible set of \emph{non-input} nodes. If there is a matching in which all of these non-input nodes are matched, then the graph is controllable. Define a set $\mathcal{I}$ by $$A \in \mathcal{I} \Leftrightarrow \mbox{There exists a matching where A is matched}.$$  The following result maps controllability to a matroid constraint.
\begin{theorem}[\cite{clark2012controllability}]
\label{prop:SC_matroid}
The tuple $(V,\mathcal{I})$ defines a transversal matroid.
\end{theorem}

Controllability can therefore be expressed as a matroid constraint $V \setminus S \in \mathcal{I}$. Many of the properties of this matroid have  physical interpretations. The bases of the matroid correspond to the minimum-size input sets. For any set $T \subseteq V$, the rank function $r(T)$ is equal to the maximum number of nodes in $T$ that are matched, and hence is equivalent to the number of nodes that are controllable when the input set is $S = V \setminus T$. 

A related problem to selecting a minimum-size set of input nodes for controllability is determining how effective a given set of input nodes is at controlling a graph. One graph-based controllability metric, denoted as the \emph{graph controllability index (GCI)}, is defined by~\cite{clark2012controllability} $$GCI(S) = \max{\{|V^{\prime}| : \mbox{Graph $(V^{\prime}, E(V^{\prime}))$ is controllable from $S$}\}}$$ where $E(V^{\prime})$ is the set of edges in $E$ that are between nodes in $V^{\prime}$. As an example, the GCI of the graph shown in Figure \ref{fig:controllability_matching} is 6, since all non-input nodes are matched under a maximal matching. If $S = \{n_{5}\}$, then the maximum-cardinality matching that can be obtained is $4$, implying that a total of five nodes (one input and four non-input nodes) are controllable. 

In order to compute the GCI, observe that the set of nodes that are controllable can be decomposed into the set of input nodes and the set of controllable non-input nodes. The set of input nodes has cardinality $|S|$ by definition. The set of controllable non-input nodes, by the preceding discussion, has cardinality $r(V \setminus S)$. Hence the graph controllability index can be written as $$GCI(S) = r(V \setminus S) + |S|,$$ which is the sum of a matroid rank function and the cardinality function and hence is monotone increasing and submodular. 


\subsection{Minimizing Controller Energy}
\label{subsec:energy}
Controllability refers to the ability of the controller to steer the network states to any desired values in a finite time by providing arbitrary input signals. In practice, however, an arbitrary input signal may require high levels of energy, making control from a given input set infeasible even if the controllability condition is satisfied. The minimum control effort problem for a given input set is formulated as
\begin{equation}
\label{eq:min_energy}
\begin{array}{ll}
\mbox{minimize} & \int_{t_{0}}^{t_{1}}{u(t)^{T}u(t) \ dt} \\
u(t): t \in [t_{0},t_{1}] & \\
\mbox{s.t.} & \dot{x}(t) = Ax(t) + Bu(t), \ t \in (t_{0},t_{1}] \\
 & x(t_{0}) = x_{0}, \ x(t_{1}) = x_{1}
 \end{array}
\end{equation}
The solution to this optimization problem is characterized by the controllability Gramian $W_{S}$, and results in a minimum energy given by $$(x_{1}-e^{A(t_{1}-t_{0})})^{T}\Gamma(t_{0},t_{1})^{-1}(x_{1}-e^{A(t_{1}-t_{0})}x_{0})$$ where $\Gamma(t_{0},t_{1})$ is the controllability matrix. The impact of the choice of input nodes on the controllability Gramian can be seen for the case where the $B$ matrix is diagonal, so that each incoming signal impacts exactly one input node. In this case, the matrix $\Gamma$ can be written as $\sum_{i \in S}{\Gamma_{i}}$, where $\Gamma_{i} = \int_{t_{0}}^{t_{1}}{e^{At}\delta_{i}e^{A^{T}t} \ dt}$ and $\delta_{i}$ is a matrix with a $1$ in the $(i,i)$-th entry and zeros elsewhere. The problem of selecting a minimum-size set of inputs to ensure that the controller energy is below a desired value $R$ can then be formulated as 
\begin{equation}
\label{eq:energy_form} 
\begin{array}{ll}
\mbox{minimize} & |S| \\
\mbox{s.t.} & v^{T}\Gamma^{-1}v \leq R
\end{array}
\end{equation}
where $v = (x_{1}-e^{A(t_{1}-t_{0})}x_{0})$. A submodular function that is arbitrarily close to the constraint in (\ref{eq:energy_form}) is given by $$f_{\epsilon}(S) \triangleq v^{T}(\Gamma + \epsilon I)^{-1}v + \epsilon\sum_{i=1}^{n-1}{\overline{v}_{i}^{T}(\Gamma + \epsilon^{2}I)^{-1}\overline{v}_{i}},$$ where $\overline{v}_{1},\ldots,\overline{v}_{n-1}$ are an orthonormal basis for the null space of $v$. 

\begin{theorem}[\cite{tzoumas2015minimal}]
\label{theorem:energy_supermodular}
For any $\epsilon > 0$, the function $f_{\epsilon}(S)$ is supermodular as a function of $S$.
\end{theorem}

The results of this section imply that, for a variety of systems, selecting a set of input nodes to satisfy controllability can be formulated as a submodular optimization problem, implying the existence of computationally efficient and provably optimal input selection algorithms for controllability. Application domains include selecting a subset of leaders in a leader-follower formation network in order to ensure that any specified trajectory can be followed; choosing a subset of genes to ensure that a cell can be steered to a desired final state; and selecting a set of generators to control in order to stabilize a power system.
\section{Putting It Together: Performance and Controllability}
\label{sec:perf_controll}

\begin{figure}[!ht]
\centering
\includegraphics[width=5in]{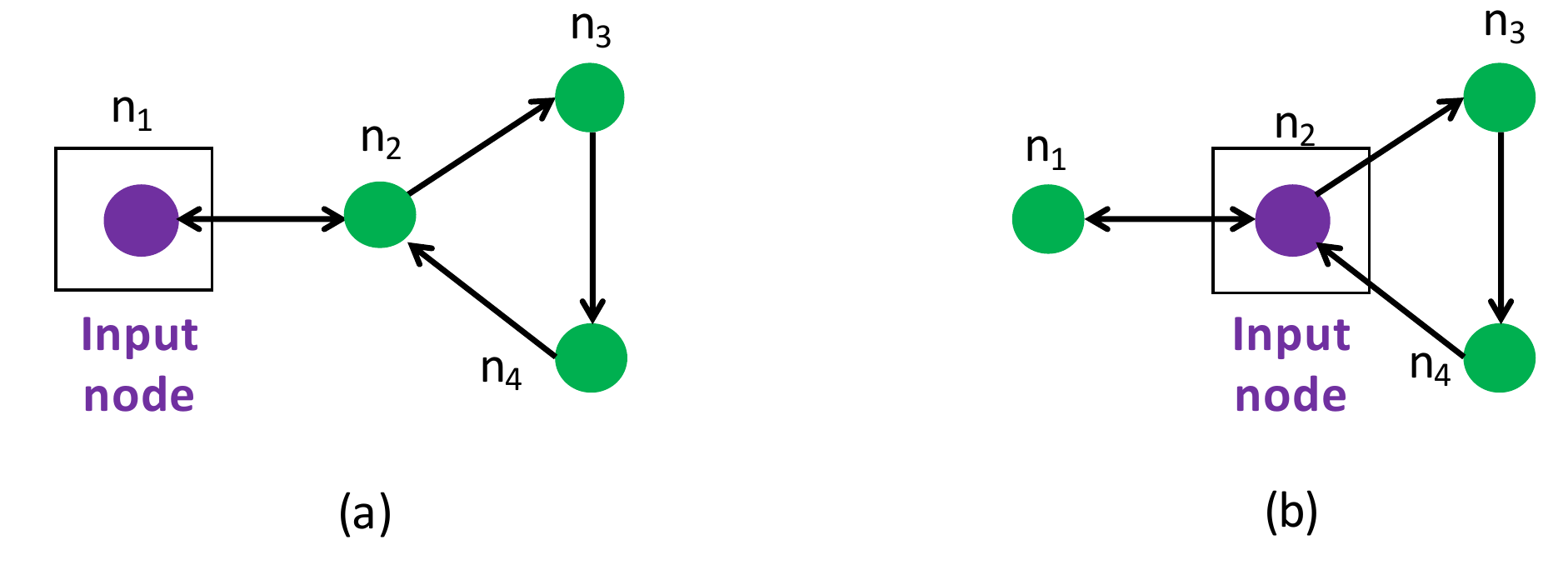}
\caption{Comparison of two possible input nodes. (a) The chosen input node satisfies controllability but provides poor performance due to its distance from the non-input nodes. (b) The input node $n_{2}$ is centrally located but does not satisfy controllability, since there is a dilation $A \subseteq V \setminus \{n_{2}\}$ with $A = \{n_{1},n_{3}\}$ and $N(A) = \{n_{2}\}$.}
\label{fig:comparison}
\end{figure}

Consider the two networks in Figure \ref{fig:comparison}.
The network on the left can be shown to satisfy controllability (see  the section``Submodularity and Controllability''), however, the chosen input node is distant from the remaining network nodes, and hence suboptimal for performance metrics such as smooth convergence, which rely on inputs reaching the remaining non-input nodes in a timely fashion. On the other hand, in the network on the right, the input node is centrally located but does not guarantee controllability. The potential conflicts between different design requirements motivates the development of an analytical framework for joint input selection based on performance and controllability.

The problem of selecting a set of up to $k$ input nodes to ensure structural controllability while maximizing a performance metric is given by
\begin{equation}
\label{eq:joint_perf_controll}
\begin{array}{ll}
\mbox{maximize} & f(S) \\
\mbox{s.t.} & \mbox{Network controllable from $S$} \\
 & |S| \leq k
\end{array}
\end{equation}

The following theorem leads to a submodular approach to joint performance and controllability.
\begin{theorem}
\label{theorem:joint_perf_controll}
If the function $f(S)$ is monotone, then there exists a matroid $\mathcal{M}$ such that Problem (\ref{eq:joint_perf_controll})  is equivalent to $\max{\{f(S) : S \in \mathcal{M}\}}$.
\end{theorem}


If the function $f(S)$ is submodular, then this problem is  submodular maximization subject to a matroid constraint. A modified version of the greedy algorithm suffices to approximate this problem with provable optimality guarantees:
\begin{enumerate}
\item Initialize $S = \emptyset$.
\item If $|S| = k$, return $S$. Else go to 3.
\item Select $v$ satisfying $(S \cup \{v\}) \in \mathcal{M}$ and $v$ maximizes $f(S \cup \{v\})$.
\item Set $S \leftarrow (S \cup \{v\})$. Go to 2.
\end{enumerate}
The greedy algorithm is guaranteed to achieve an optimality bound of $1/2$~\cite{fischer1978analysis}. This optimality bound can be improved to $(1-1/e)$ through the continuous greedy algorithm~\cite{calinescu2011maximizing}; the complexity of the algorithm, however, precludes its use on large-scale networks.

An implicit assumption in (\ref{eq:joint_perf_controll}) is that there exists a set of input nodes with cardinality $k$ that are sufficient for controllability. This condition can be relaxed through the graph controllability index $GCI(S)$, so that the optimization problem becomes $\max{\{f(S) + \lambda GCI(S) : |S| \leq k\}}$ where $\lambda$ is a nonnegative constant. 

Finally, for systems with multiple performance and controllability constraints, the constraints can be combined as
\begin{equation}
\label{eq:joint_constraints}
\begin{array}{ll}
\mbox{minimize} & |S| \\
\mbox{s.t.} & f_{1}(S) \leq \alpha_{1} \\
 & \vdots \\
 & f_{m}(S) \leq \alpha_{m}
\end{array}
\Leftrightarrow
\begin{array}{ll}
\mbox{minimize} & |S| \\
\mbox{s.t.} & \sum_{i=1}^{m}{\min{\{f_{i}(S), \alpha_{i}\}}} \geq \sum_{i=1}^{m}{\alpha_{i}}
\end{array}
\end{equation}
and hence inputs can be selected jointly while maintaining provable optimality guarantees.
\section{Numerical Studies}
\label{sec:numerical}
In order to illustrate the potential benefits of the submodular optimization approach to control of networked systems, numerical studies of input selection for robustness to noise, smooth convergence, and controllability are presented. For each problem, the following algorithms were compared: (a) the submodular optimization approach, (b) selection of high-degree nodes to act as inputs, (c) selection of average-degree nodes, and (d) random input selection. Each numerical study was averaged over $50$ independent trials.

\begin{figure}[!ht]
\centering
\includegraphics[width=5in]{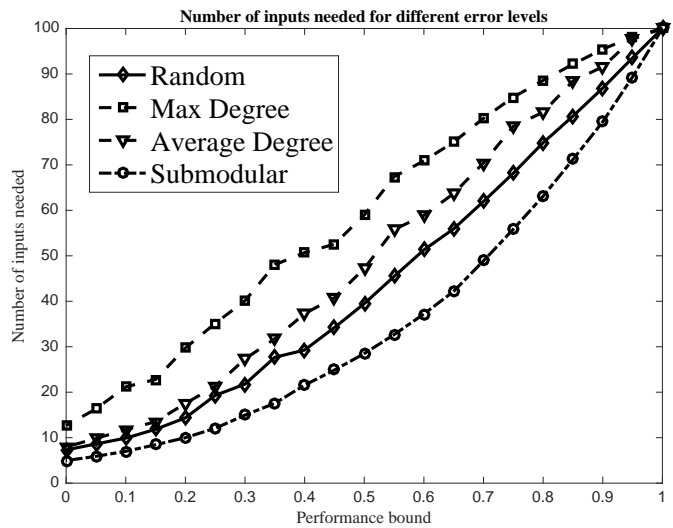}
\caption{Numerical study of input selection for robustness to noise in a network of $100$ nodes. The submodular optimization approach requires fewer inputs to achieve a desired bound on the error due to noise than the degree-based heuristics.}
\label{fig:robustness_sim}
\end{figure}

Robustness to noise was evaluated in a network of $n=100$ nodes. The network topology was generated according to a geometric model, in which nodes are deployed uniformly at random over a square region with width $w=1000$m and an undirected edge exists between nodes $i$ and $j$ if their positions are within $300$m of each other. In this scenario, the submodular approach requires 10-20 fewer inputs to achieve a desired error bound than the random and average degree heuristics, and 20-40 fewer inputs than the maximum-degree input selection.

\begin{figure}[!ht]
\centering
\includegraphics[width=5in]{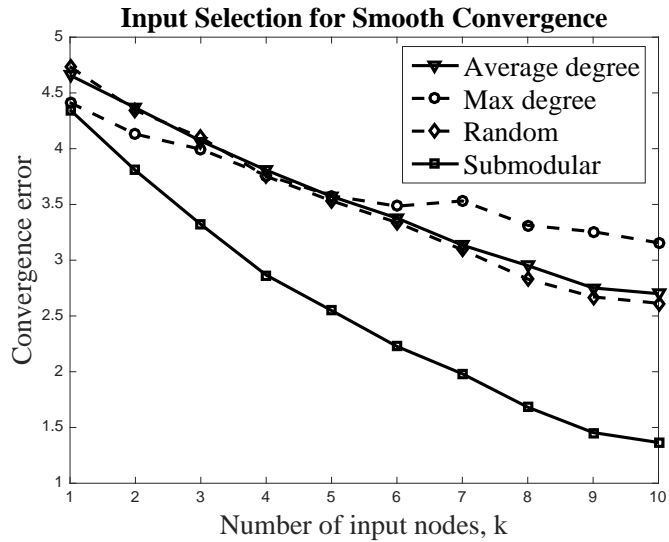}
\caption{Input selection for smooth convergence in a network with $100$ nodes. The submodular approach led to less than half of the convergence error of the degree-based and random heuristics. The diminishing returns property of the convergence error can also be seen in each of the three curves.}
\label{fig:convergence_sim}
\end{figure}

The numerical study of smooth convergence is shown in Figure \ref{fig:convergence_sim}, for a geometric graph with $n=100$ nodes, width $1400$m, $r=250$m, and $p=2$. The convergence error arising from the submodular optimization algorithm is less than half of the convergence error from random or degree-based input selection,  especially  as the number of input nodes grows. For number of input nodes $k=9$ and $k=10$, the ``diminishing returns'' property of the convergence error can be observed, as the impact of the tenth input node is reduced compared to previous inputs. 

\begin{figure}[!ht]
\centering
\includegraphics[width=5in]{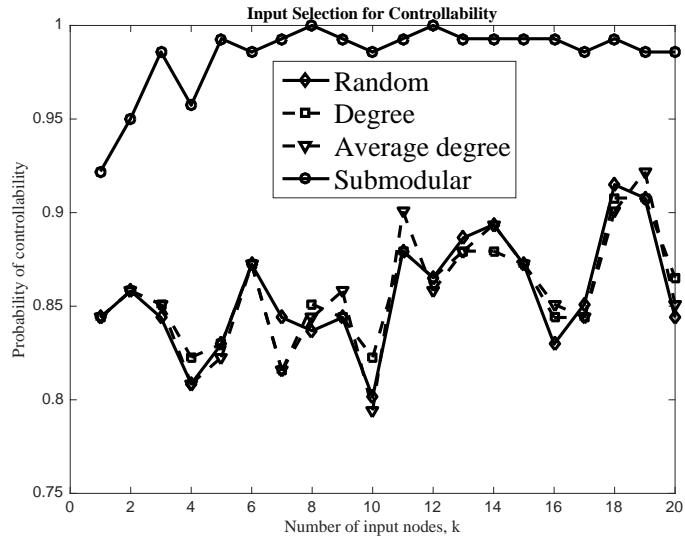}
\caption{Input selection for controllability in a network of $70$ nodes. Each data point represents the probability that the input selection method returned a set of $k$ input nodes guaranteeing controllability. The submodular approach ensures controllability in over $95\%$ of cases for $k \geq 2$, while the other heuristics only ensured controllability in $80-90\%$ of cases.}
\label{fig:controllability_sim}
\end{figure}

Numerical results of input selection for controllability are shown in Figure \ref{fig:controllability_sim}. This simulation considered an Erdos-Renyi random graph, defined as a graph of $n=70$ nodes where an edge exists between nodes $i$ and $j$ with probability $q=0.07$. The submodular approach is based on joint optimization of controllability and convergence error. The numerical results show that submodular optimization approach selects an input set that is guaranteed to satisfy controllability in 95$\%$ of the randomly generated network topologies, while the degree-based and random selection algorithms only ensure controllability in $80-90\%$  of generated network topologies. The submodular approach failed to ensure controllability during $100\%$ of trials because, for some randomly generated networks, it was not possible to select a set of $k$ input nodes that satisfy controllability.
\section{Open Problems}
\label{sec:open}
Submodularity has received attention in the control and dynamical systems community only recently, leaving extensive work to be done to generalize the preliminary approaches described in this article to new systems, operating conditions, and application domains. A sampling of the open problems in submodularity for control of networked systems is given in the following section.

\subsection{Nonlinear Network Dynamics}
\label{subsec:nonlinear}
Networked systems exhibit complex nonlinear interactions between nodes, and can sometimes be only loosely approximated using linear models. Currently, techniques for guaranteeing performance and controllability of such nonlinear systems are in the early stages. 

An important challenge in control of nonlinear systems is ensuring stability of the system. These stability guarantees are typically provided using Lyapunov methods; at present, submodular structures for energy-based and Lyapunov methods have yet to be investigated. In the area of submodularity for performance of nonlinear systems, the connection between nonlinear systems and random walk dynamics requires additional study in order to extend the submodular approach to smooth convergence and robustness to noise to a broader class of systems.

\subsection{Application-Driven Methods: Power Systems}
\label{subsec:power}
The power grid is a classic example of a nonlinear networked dynamical system, which must be controlled to provide guarantees on stability, reliability, and availability. Growing energy demand and integration of unpredictable renewable energy sources are bringing power systems closer to their capacity limits, posing new challenges for power system control. At the same time, the deployment of communication, monitoring, and real-time control infrastructures comprising the smart grid promise to create new opportunities for effective control.

Submodular optimization approaches have the potential to improve power system stability by addressing discrete design problems that arise in power system control. Preliminary work has investigated the use of submodularity in voltage control \cite{liu2016voltage}. Voltage instability is caused when the reactive power supplied is inadequate to meet reactive power demand at one or more buses, and is typically mitigated by switching on capacitor or reactor banks, which inject or withdraw reactive power. 

 Selecting a subset of devices to inject reactive power is inherently a combinatorial approach. Under a linearized model of the  system dynamics, preliminary results suggest that a submodular approach to selecting the devices could reduce the computational complexity while maintaining system performance and stability. The linearized assumption, however, breaks down as the system approaches unstable or critical points, creating a need for new techniques that provide guarantees on the underlying nonlinear dynamics.
 
 \subsection{Uncertain and Time-Varying Networks}
 \label{subsec:uncertain}
 The topologies of networked systems evolve over time. Most input selection techniques assume either a static network topology, or a topology that varies according to a known deterministic or probabilistic model. Input selection algorithms for systems with arbitrary time-varying topologies are an open research area due to the added complexity and difficulty of obtaining provable guarantees on such systems.
 
 One approach that has been studied is predicting future network topologies using experts or multi-armed bandit algorithms, and selecting time-varying input sets accordingly. These algorithms lead to provable ``no-regret'' optimality bounds~\cite{clark2014minimizing,clark2014supermodular}. There is substantial room for improvement, however, by incorporating additional information on the topology dynamics (e.g., when topology changes are induced by changes in the network states themselves, as in state-dependent graphs).
 
 \subsection{Distributed and Online Algorithms}
 \label{subsec:distributed}
 Self-organization is a basic principle of networked systems from biological networks to smart transportation. From an engineering perspective, self-organized approaches have advantages of scalability and robustness compared to top-down centralized design. These considerations motivate the development of adaptive and distributed algorithms for input selection.
 
While centralized submodular optimization techniques have been studied leading to efficient approximation algorithms, scalable methodologies for distributed submodular optimization are currently in the early stages. This is especially true in distributed networks with limited communication and sensing capabilities. While methods have been developed towards submodular optimization using local search heuristics under cardinality and matroid constraints~\cite{clark2014distributed}, open research challenges remain in providing the same optimality guarantees as centralized algorithms and reducing the communication and computation complexity.

\section{Conclusions}
\label{sec:conclusion}
This article presented submodular optimization methods for selecting input nodes in networked systems. Submodularity is a diminishing returns property of set functions that leads to efficient algorithms with provable optimality bounds for otherwise intractable combinatorial optimization problems. Submodular optimization techniques were presented for two classes of input selection problem, namely, input selection for performance and input selection for controllability, as well as joint selection based on both of these criteria.

In the area of input selection for performance, the submodular approach was developed for robustness to noise and smooth convergence to a desired state. It was shown that, for linear consensus dynamics, the mean-square error in steady-state is a supermodular function of the input set. The  error experienced in Kalman filtering was also studied and shown to be a supermodular function of the input set. Smooth convergence was discussed by bounding the norm of the deviation of the node states from the convex hull of the input nodes, which was then shown to be a supermodular function. In all cases, the supermodularity property was derived from connections between the performance metrics and the statistics of random walks on the network.

Input selection for controllability was investigated for systems with known, fixed parameters, as well as structured systems with unknown parameters. In the latter case, a submodular framework was developed by mapping structural controllability criteria to graph matchings. Finally, energy-related metrics, such as the minimum energy required to drive the system to a desired state, were considered. 

Submodular algorithms for networked control are still in their early stages, with many open questions including extensions to nonlinear dynamics, domain-specific challenges in applications such as power systems, and moving from centralized to distributed approaches. New insights in any of these areas have the potential to not only improve the stability and performance of dynamical systems, but also have broader applications in machine learning and optimization. 
\newpage
\bibliographystyle{unsrt}
\bibliography{CSM_2016}

\eject

   \end{document}